*Heaven's Light is Our Guide*

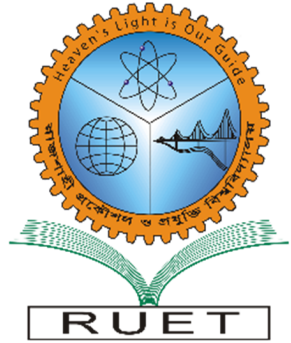

# DEPARTMENT OF COMPUTER SCIENCE & ENGINEERING

## Rajshahi University of Engineering & Technology, Bangladesh

## Prognostic Biomarker Identification for Pancreatic Cancer by Analyzing Multiple mRNA Microarray and microRNA Expression Datasets

**Author**
Md. Azmain Yakin Srizon
Roll No. 143005
Department of Computer Science & Engineering
Rajshahi University of Engineering & Technology

**Supervised by**
Dr. Md. Al Mehedi Hasan
Professor
Department of Computer Science & Engineering
Rajshahi University of Engineering & Technology

# ACKNOWLEDGEMENT

At first, I would like to thank the Almighty Allah for giving me the opportunity and enthusiasm along the way for the completion of my thesis work.

I would like to express my sincere appreciation, gratitude, and respect to my supervisor **Dr. Md. Al Mehedi Hasan,** Professor, Department of Computer Science and Engineering, Rajshahi University of Engineering and Technology, Rajshahi. Throughout the year he has not only given me technical guidelines, advice and necessary documents to complete the work but also, he has given me continuous encouragement, advice, helps and sympathetic co-operation whenever he deemed necessary. His continuous support was the most successful tool that helped me to achieve my result. Whenever I was stuck in any complex problems or situation, he was there for me at any time of the day. Without his sincere care, this work not has been materialized in the final form that it is now at the present.

I want to express gratitude and thanks to **Professor Dr. Boshir Ahmed**, Head, Department of Computer Science & Engineering, Rajshahi University of Engineering & Technology, for his extending helps in various ways from his department.

I am also grateful to all the respective teachers of Computer Science and Engineering, Rajshahi University of Engineering and Technology, Rajshahi for good & valuable suggestions and inspirations from time to time.

Finally, I convey my thanks to my parents, sister, friends, and well-wishers for their constant inspirations and many helpful aids throughout this work.

August 2019　　　　　　　　　　　　　　　　　　　　　　　Md. Azmain Yakin Srizon
RUET, Rajshahi

*Heaven's Light is Our Guide*

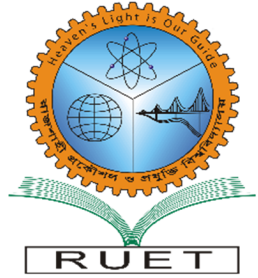

**DEPARTMENT OF COMPUTER SCIENCE & ENGINEERING**

Rajshahi University of Engineering & Technology, Bangladesh

# *CERTIFICATE*

*This is to certify that this thesis report entitled* ***"Prognostic Biomarker Identification for Pancreatic Cancer by Analyzing Multiple mRNA Microarray and microRNA Expression Datasets"*** *submitted by* ***Md. Azmain Yakin Srizon, Roll:143005*** *in partial fulfillment of the requirement for the award of the degree of Bachelor of Science in Computer Science & Engineering of Rajshahi University of Engineering & Technology, Bangladesh is a record of the candidate own work carried out by him under my supervision. This thesis has not been submitted for the award of any other degree*

| Supervisor | External Examiner |
|---|---|
| -------------------------------------------- | -------------------------------------------- |
| **Dr. Md. Al Mehedi Hasan** | **Julia Rahman** |
| Professor | Assistant Professor |
| Department of Computer Science & Engineering | Department of Computer Science & Engineering |
| Rajshahi University of Engineering & Technology | Rajshahi University of Engineering & Technology |
| Rajshahi-6204 | Rajshahi-6204 |

# ABSTRACT

Possessing the five-year durability rate of nearly 5%, currently, the fourth leading cause for cancer-related deaths is pancreatic cancer. Previously, several works have resolved that early diagnosis performs a meaningful function in enhancing the durability rate and diverse online tools have been utilized to distinguish prognostic biomarker which is a lengthy process. We believe that the statistical feature selection method can produce a better and faster result here. To authenticate our statement, we picked three different mRNA microarray (GSE15471, GSE28735, and GSE16515) and a microRNA (GSE41372) dataset for identification of differentially expressed genes (DEGs) and differentially expressed microRNAs (DEMs). By adopting some feature selecting methods, 178 DEGs and 16 DEMs were elected. After identifying target genes of DEMs, we selected two DEGs (ECT2 and NRP2) which were also identified among DEMs target genes. Moreover, overall durability report established that ECT2 and NRP2 were associated with poor overall survival. Hence, we concluded that for pancreatic cancer, statistical feature selection approaches certainly perform better for biomarker identification than pre-defined online programs, and here, ECT2 and NRP2 can act as possible prognostic biomarkers. All the resources, programs and snippets of our literature can be discovered at https://github.com/Srizon143005/PancreaticCancerBiomarkers.

# CONTENTS

# LIST OF TABLES

# LIST OF FIGURES

# CHAPTER 1
# Introduction

*Introduction*

*Motivation*

*Problem Statements*

*Research Objectives*

*Research Contribution*

*Organization of Thesis*

*Conclusion*

## 1.1 Introduction

Noncommunicable diseases (NCDs) are now accountable for the majority of global mortality, [1] and cancer is supposed to rank as the foremost reason of mortality and the single most significant obstacle to improving life anticipation in every country of the world in the 21st century. According to assessments from the World Health Organization (WHO) in 2015, cancer is the first or second leading reason of death before age 70 years in 91 of 172 countries, and it stands third or fourth in further 22 countries (Figure 1.1) [2]. Cancer frequency and death are swiftly expanding worldwide. The causes are complicated but consider both aging and increase of the population, as well as variations in the prevalence and arrangement of the main risk factors for cancer, some of which are correlated with socioeconomic improvement [3-4].

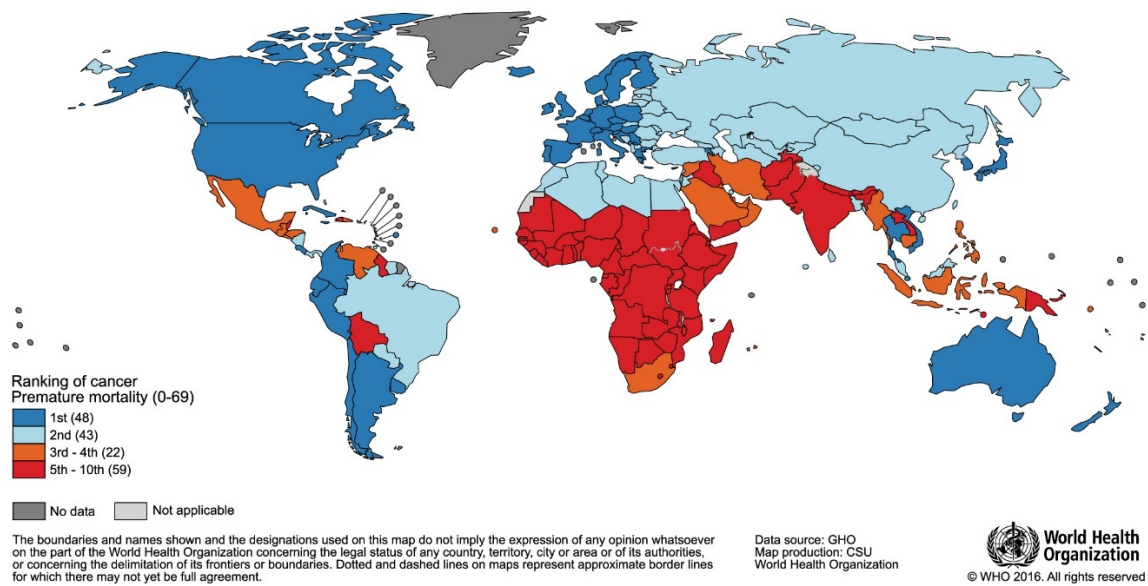

Figure 1.1: Global Map Presenting the National Ranking of Cancer as a Cause of Death at Ages Below 70 Years in 2015 [2]

Pancreatic carcinoma is amongst the diverse most malignant types of cancer with a five-year survival rate of only 2-8% [5]. Rapid progression of the disease, inadequate response to palliative regimens and diagnosis at advanced stage are the main causes of poor survival rate in pancreatic cancer. Because of having no effective method of diagnosis at the initial stage and as maximum pancreatic cases are initially diagnosed at an advanced stage [6-7], molecular biomarkers may become beneficial during monitoring, prognosis and diagnosis. Having vast openly accessible data and advanced online programs for analysis, it is quite evident for the scientific community to find molecular biomarkers as it is an important diagnostic tool [8].

The overwhelming majority (>80%) of victims with pancreatic cancer present with locally advanced disease or distant metastases, rendering cancer surgically inoperable [9]. Hence, the discovery of pancreatic cancer at an initial, and hence possibly resectable stage, allows the best hope for a cure. Unfortunately, tumor markers that are currently employed for the discovery of pancreatic cancer in clinical practice lack the sensitivity and specificity needed to detect potentially curable lesions. For example, serum CA-19-9 is considered the best test for pancreatic cancer [10]; however, CA-19-9 can also be raised in a myriad of non-neoplastic infirmities such as severe and persistent pancreatitis, hepatitis, and biliary obstruction, greatly decreasing its specificity [11]. Besides, victims with particular blood group antigen types do not expose the CA-19-9 antigen. Consequently, the sensitivity of CA- 19-9 approaches ~80%, restricting its use for screening objectives particularly for the diagnosis of localized, resectable pancreatic cancers [12]. The dismal prognosis and late presentation of pancreatic cancer in most people indicate the necessity for producing an advanced early disclosure strategy. This is particularly true as progress in our understanding of the genetics of pancreatic cancer has assisted identify families and individuals who harbor a derived predisposition to this disease [13]. Enhanced early diagnosis procedures in such at-risk victims could apparently save lives [14].

## 1.2    Motivation

Mortality from pancreatic ductal adenocarcinoma, also known as pancreatic cancer, rank fourth among cancer-related deaths in the United States. In 2016, 53,070 victims were diagnosed with pancreatic carcinoma and it caused 71,780 estimated deaths in United States [15]. Recent studies revealed pancreatic cancer is one of the ten most frequent cancers in both male and female that caused 79,400 losses in China [16] along with 330,400 losses globally [17]. In Europe, pancreatic cancer holds a five-year survival rate of roughly 7% [15]. The average survival time after being diagnosed is 6 months for pancreatic cancer [18] and by 2030 it is anticipated to be at second spot following lung cancer [19].

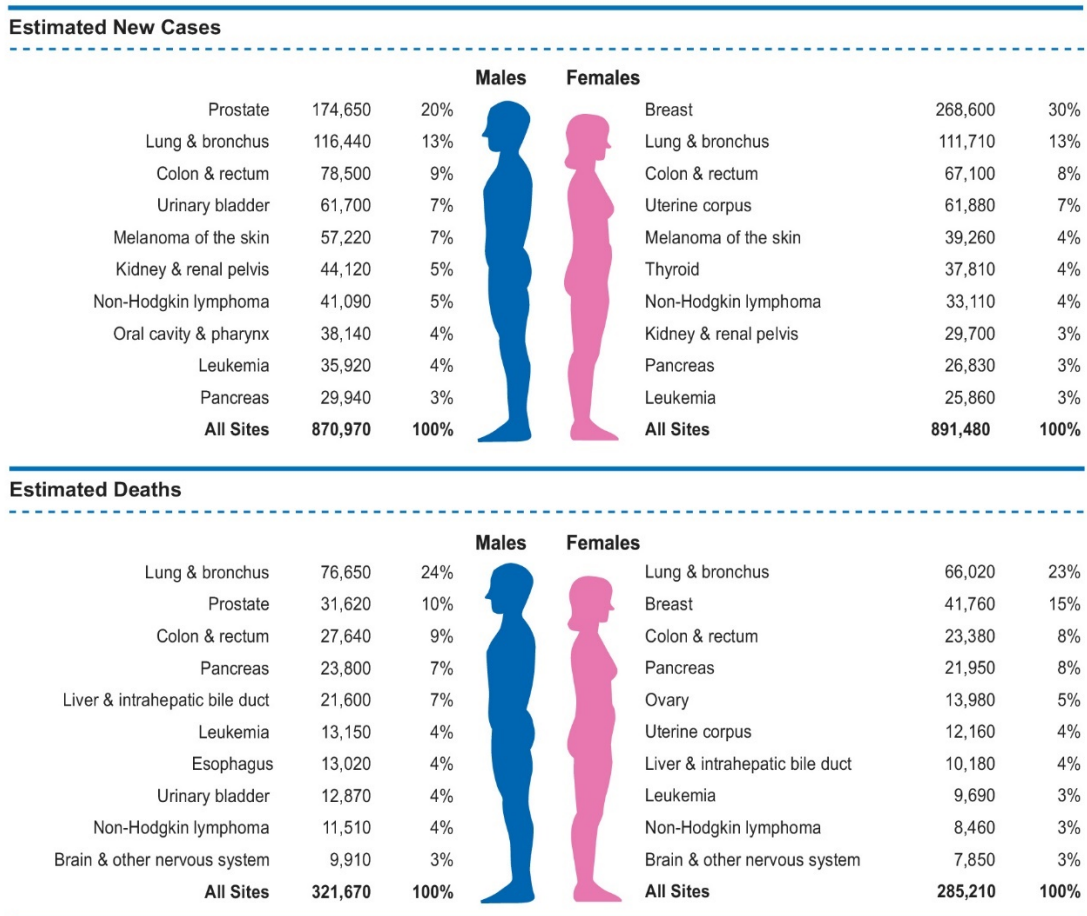

Figure 1.2: Ten Leading Cancer Types for the Estimated New Cancer Cases and Deaths by Sex, United States, 2019 [20]

Cancer statistics proposed that in USA, approximately 56,770 new cases of pancreatic cancer will be introduced in 2019 and among them estimated death count will be 45,750 [20]. Figure 1.2 displays that pancreatic cancer will hold its fourth place as per estimated deaths in 2019 as well for both male and female [20]. Global Cancer Statistics 2018 revealed that because of its poor prognosis, with almost as many deaths (d = 432,000) as cases (n = 459,000), pancreatic cancer is the seventh leading cause of cancer death in both males and females possessing worst survival rate [2]. Rates are 3-fold to 4-fold higher in higher HDI countries, with incidence rates highest in Europe, North America, and Australia/New Zealand (Figure 1.3) [2]. In the 28 countries of the European Union, given that rates are rather stable relative to diminishing rates of breast cancer, it has been predicted that pancreatic cancer will outdo breast cancer as the third preeminent cause of cancer-related death in the future. [21]

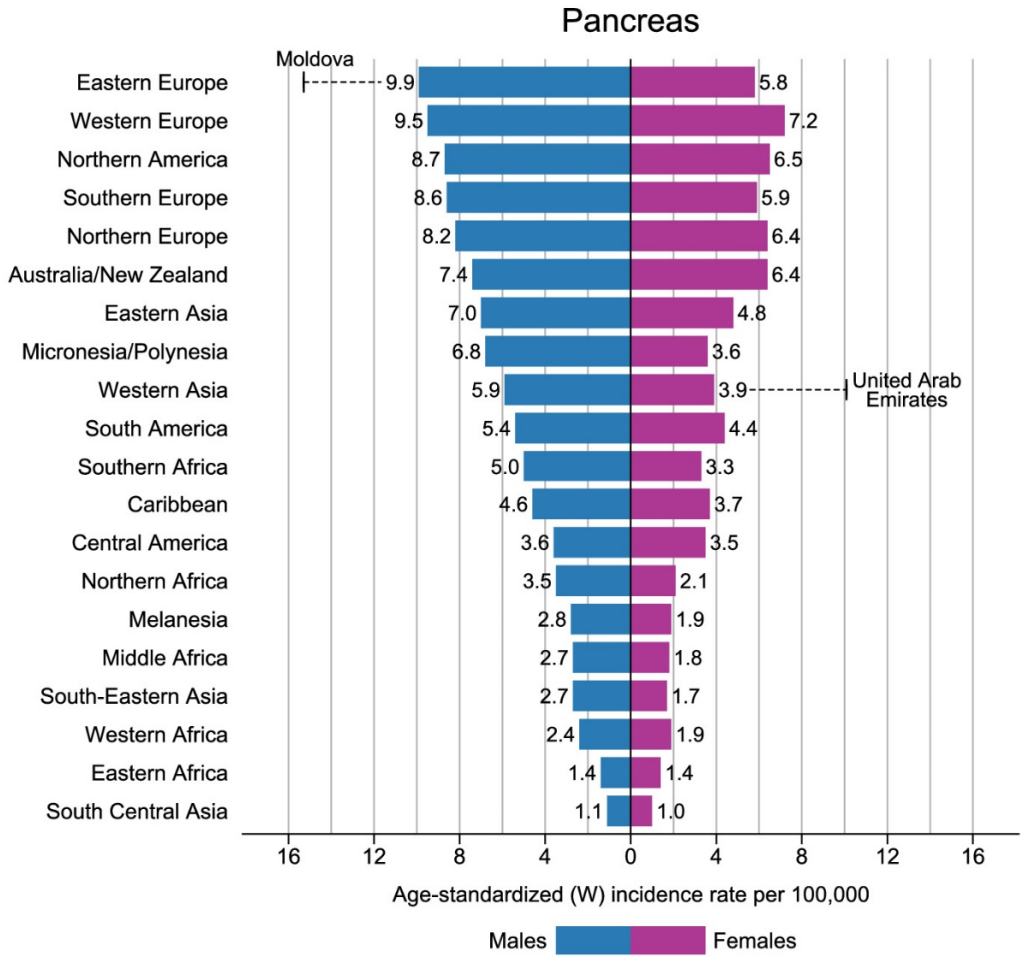

Figure 1.3: Bar Chart of Region-Specific Incidence Age-Standardized Rates by Sex for Pancreatic Cancer in 2018 [2]

Pancreatic cancer is more prevalent in elderly people than in younger persons, and less than 20% of patients present with localized, possibly curable tumors. The overall 5-year survival rate among patients with pancreatic cancer is <5% [22-23]. The reasons of pancreatic cancer remain anonymous. Various environmental circumstances have been compromised, but evidence of a causative role exists only for tobacco consumpyion. The risk of pancreatic cancer in smokers is 2.5 to 3.6 times that in nonsmokers; the danger rises with greater tobacco consumption and longer appearance to smoke [24]. Data are limited on the potential functions of controlled consumption of alcohol, absorption of coffee, and use of aspirin as offering factors. Several researches have revealed an enhanced incidence of pancreatic cancer among victims with a history of diabetes or chronic pancreatitis, and there is also testimony, although less convincing, that chronic cirrhosis, a high-fat, high-cholesterol diet, and previous cholecystectomy are correlated with an enhanced incidence [25-28]. More recently, an improved risk has been recognized among patients with blood type A, B, or AB as corresponded with blood type O [29].

Almost 5-10% of victims with pancreatic cancer have a family background of the disease [30]. In several victims, pancreatic cancer improves as part of a distinct cancer-predisposing syndrome for which germ-line genetic modifications are acknowledged. Besides, in some families with an enhanced risk of pancreatic cancer, a genetic rather than an environmental reason is doubted. The risk of pancreatic cancer is 57 times as high in families with four or more affected persons as in families with no affected persons [31]. The genetic bases for these correlations are anonymous, although a subgroup of such high-risk kindred carry germ-line mutations of DNA repair genes such as BRCA2 and the companion and localizer of BRCA2 (PALB2) [32-34].

## 1.3 Problem Statements

Victims with pancreatic cancer that was incidentally diagnosed while imaging examination for the independent disease have a more expanded median survival time than symptomatic patients [35]. These data imply that early exposure is the key issue for developing the prognosis of this destructive disease. Yachida *et al.* described that the time from the initiating mutation in the pancreas to increase of metastatic pancreatic cancer is almost 21 years [36]. This presents a wide window for early exposure of pancreatic cancer. Previous studies have established the crucial function of early detection in the prognosis of victims with pancreatic cancer.

Pancreatic cancer-oriented biomarker identification is vital for early diagnosis [37] and the main resources for biomarker identification are microarray data [38]. In gene expression analysis of microarray data, there are thousands of genes expressions present.

Statistical methods can be used to obtain the differentially expressed genes (DEGs). These genes, then, can be matched with the target genes of differentially expressed microRNAs that can also be identified by using statistical methods. MicroRNA expression datasets contain hundreds of microRNAs, therefore, identifying differentially expressed microRNA can be significant for biomarker identification.

Previous works showed that by using online tools and analyzing the selected genes, the prognostic biomarker can be chosen [39]. But better results may be obtained by statistical feature selection approaches. Hence, in this research, we selected prognosis of pancreatic cancer as the problem domain. The main cause was the lack of treatment in an advanced stage of pancreatic cancer. In the past several years, the diagnosis methods have not been improved and the survival rate of pancreatic cancer is moving down. In this scenario, working with prognosis or early detection seems more significant to save lives.

## 1.4   Research Objectives

The main objective of this research is to identify the prognostic biomarkers for pancreatic cancer. To obtain prognostic biomarkers of pancreatic cancer, the first step we require to apply is dimension reduction. However, we can't apply feature extraction techniques for this research as biomarkers have to be genes explicitly since feature extraction techniques combine the features when applied. Hence, in this study, we have used different feature selection approaches. Therefore, the objectives of our research can be shown in the following order:

- ❖ Selection of the best feature selection method for identifying differentially expressed genes and microRNAs
- ❖ Identification of the prognostic biomarkers utilizing target genes of microRNAs
- ❖ Validation of prognostic biomarkers by using overall survival analysis

## 1.5   Research Contribution

The main contribution of the research is the identification of the prognostic biomarkers. Previous studies revealed that by using online tools prognostic biomarkers can be identified. The most recent study suggested that the prognostic biomarkers can be ECT2, NR5A2, NRP2 and TGFBI. In our research, we have applied statistical feature selection approaches and several validations and analytical tools to identify the prognostic biomarkers and the outcome was satisfactory. We have identified two prognostic biomarkers: ECT2 and NRP2. The contribution of our research should assist the researchers to clinically analyze the importance and biological impact of ECT2 and NRP2 on the prognosis or the early-stage diagnosis of pancreatic cancer.

## 1.6   Organization of Thesis

**Chapter 2 - Pancreatic Ductal Adenocarcinoma**

This chapter introduces the biology, clinical representation, diagnosis, staging, management of early disease, incidence, mortality and survival of pancreatic cancer.

**Chapter 3 - Background Study and Literature Review**

This chapter discusses the biological terms required to understand for our research and describes the previous researches from diverse perspectives.

**Chapter 4 - Materials and Dataset Processing**

This chapter describes the materials and datasets associated with this research and discusses data processing keywords, for example, p-value, upregulation, downregulation, fold-change, log transformation etc.

**Chapter 5 - Methodology**

This chapter introduces different feature selection approaches: Student's T-Test, Kruskal-Wallis Test, Wilcoxon-Mann-Whitney Test and discusses p-adjustment and different p-adjustment methods along with miRecords and OncoLnc.

**Chapter 6 - Implementation and Experimental Analysis**

This chapter displays the step by step implementation and all required tables and figures with proper description for our research.

**Chapter 7 - Conclusion and Future Scopes**

This chapter concludes with the findings of our research mentioning the aspects and future scopes of this study.

## 1.7   Conclusion

In this chapter, we have attempted to present the importance of the cancer-related study. We have also introduced the significance of our research and presented why we have selected prognosis of pancreatic cancer as our research domain. Furthermore, we have discussed the research objectives, contribution and thesis organization to conclude the chapter.

# CHAPTER 2
# Pancreatic Ductal Adenocarcinoma

*Introduction*

*The Biology of Pancreatic Cancer*

*Clinical Presentation, Diagnosis and Staging*

*Staging of Pancreatic Cancer*

*Management of Early Disease*

*Incidence*

*Mortality*

*Survival*

*Conclusion*

## 2.1 Introduction

During current times, significant improvements have been accomplished in the perception of the atomic biology of Pancreatic Ductal Adenocarcinoma as well as in diagnosis, staging, and therapy in victims with early-stage tumors. However, insignificant advancement has been performed in prevention, early diagnosis, and treatment in patients with an advanced stage. This chapter interprets current advancement in the perception and supervision of pancreatic cancer.

## 2.2 The Biology of Pancreatic Cancer

Previous studies recommend that pancreatic cancer is the outcome of the consecutive accumulation of gene mutations [40]. The origin of the cancer is the ductal epithelium that emerges from premalignant lesions to completely invasive cancer. The lesion named pancreatic intraepithelial neoplasia is the stablest-characterized histologic ancestor of pancreatic carcinoma [41]. The improvement from minimally dysplastic epithelium i.e., pancreatic intraepithelial neoplasia grades 1A and 1B, to further severe dysplasia i.e., pancreatic intraepithelial neoplasia grades 2 and 3, and ultimately to invasive carcinoma is correlated with the consecutive growth of mutations that involve energizing of the KRAS2 oncogene, deactivation of the tumor-suppressor gene TP53 and CDKN2A that encodes significant inhibitor of cyclin-reliant kinase 4 (INK4A) and is removed in pancreatic cancer 4 (DPC4, likewise recognized as the SMAD species segment 4 gene [SMAD4]) [42].

These chain of events in pancreatic carcinogenesis is established by researches in genetically directed mouse designs in which targeted activation of Kras2 with simultaneous inactivation of Trp53 or Cdkn2A/Ink4A ends in the advancement of pancreatic carcinoma that is indistinguishable to the analogous human condition [43-45]. Additional less properly portrayed premalignant lesions of the pancreas encompass intrapancreatic mucinous neoplasia and mucinous cystic neoplasia [46]. Nearly every victim with sufficiently verified pancreatic cancer sustain one or more of four genetic flaws [47]. 90% of tumors possess activating modifications in the KRAS2 oncogene.

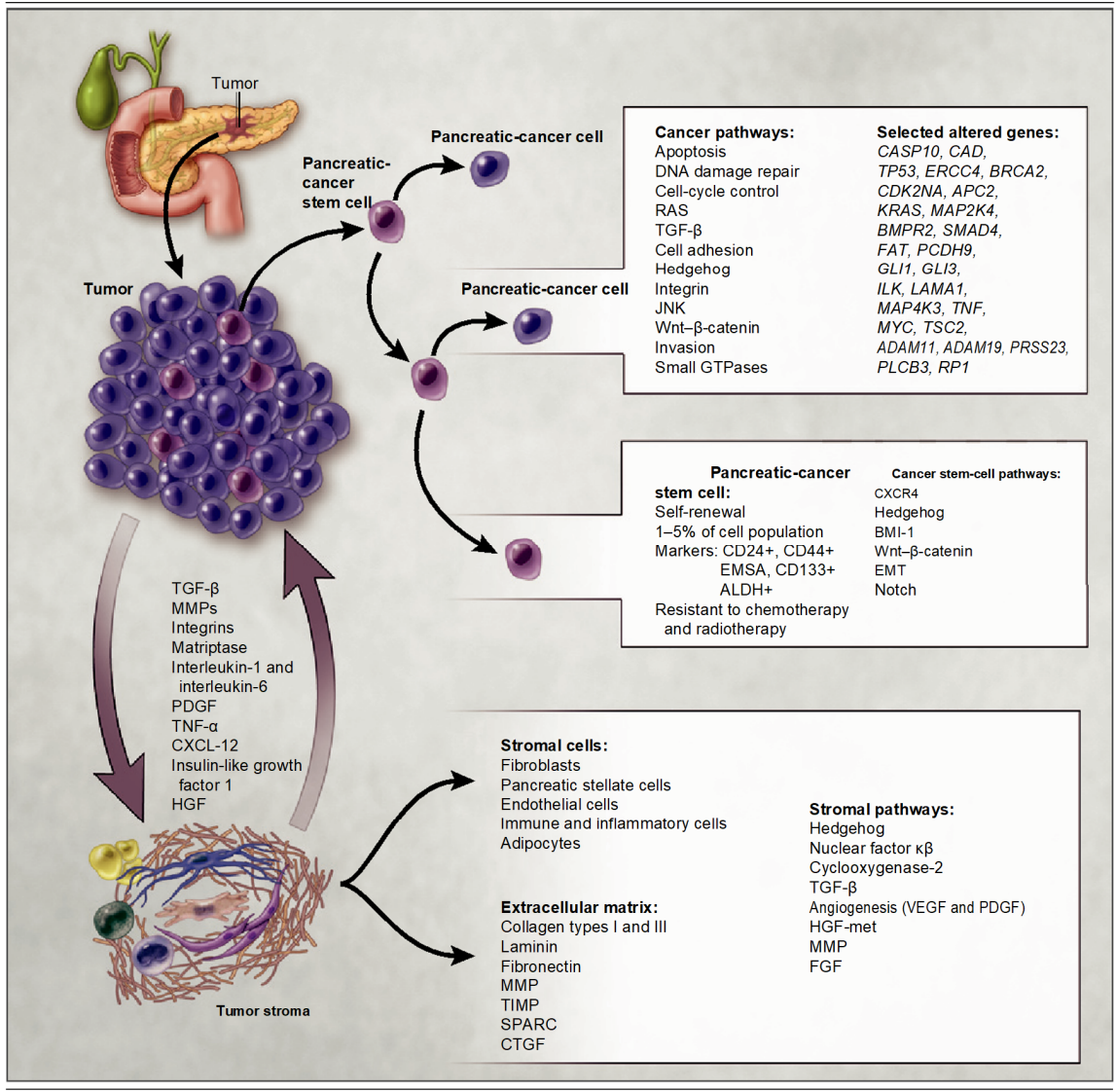

Figure 2.1: Components of Pancreatic Cancer [48]

Transcription of the mutant KRAS gene generates an irregular Ras protein which is "locked" in its activated mode, ending in anomalous activation of proliferative and durability flagging pathways. Furthermore, 95% of tumors hold inactivation of the CDKN2A gene, with the resultant sacrifice of the p16 protein (a controller of the G1–S transformation of the cell-cycle) and an identical improvement in cell reproduction. TP53 is unusual in 50-75% of tumors, authorizing cells to detour DNA erosion control checkpoints and apoptotic signs and offering to genomic vulnerability. DPC4 is suffered in roughly 50% of pancreatic cancers, ending in unusual signaling by the transforming growth factor β (TGF-β) cell-surface receptor. A current extensive genetic investigation of 24 pancreatic cancers revealed that the genetic background of pancreatic cancer is remarkably complicated and heterogeneous. [32] In such research, an aggregate of 63 genetic irregularities per tumor, essentially point modifications, were categorized as expected to be consistent. These irregularities can be adjusted in 12 operative cancer-associated pathways (Figure 2.1). Nevertheless, not all tumors have modifications in every pathway, and the fundamental modifications in every pathway seem to deviate from one tumor to another.

A feature of pancreatic cancer is the development of a compact stroma termed a desmoplastic reaction (Figure 2.1) [49]. The pancreatic stellate cells (also identified as myofibroblasts) perform a crucial function in the development and turnover of the stroma. On activation by growth factors such as TGFβ1, platelet-derived growth factor (PDGF), and fibroblast growth factor, these cells cover collagen and other ingredients of the extracellular matrix; stellate cells also resemble to be liable for the lower vascularization that is representative of pancreatic cancer [50-51]. Moreover, stellate cells coordinate the reabsorption and turnover of the stroma, essentially by the generation of matrix metalloproteinases [52]. The stroma is not just a construction block; rather, it develops a dynamic compartment that is critically linked in the process of tumor development, improvement, intrusion, and metastasis [49]. Stromal cells display various proteins such as cyclooxygenase-2, PDGF receptor, vascular endothelial growth factor, stromal cell-derived factor, chemokines, integrins, SPARC (secreted protein, acidic, cysteine-rich), and hedgehog pathway elements, amongst others, that have been correlated with a reduced prognosis and resistance to treatment. Nevertheless, these proteins may also design new therapeutic targets [53-54].

The function of angiogenesis in pancreatic cancer prevails controversial. Although initial data proposed that pancreatic cancer is angiogenesis-dependent, as are most solid tumors, treatment with angiogenesis inhibitors has deserted in patients with pancreatic cancer. A recent study in a mouse model explained that targeting the stromal hedgehog pathway improves tumor vascularization, ending in improved distribution of chemotherapeutic brokers to pancreatic tumors and higher effectiveness [55].

Besides, a subgroup of cancer cells with cancer stem-cell characteristics such as tumor induction have been recognized within the tumor [56-57]. These cells, which comprise just 1-5% of the tumor, are competent of extensive self-renewal, and through asymmetric division, they provide growth to more-differentiated organisms (Figure 2.1). Pancreatic cancer stem cells are immune to chemotherapy and radiation therapy, which may reveal why these approaches do not heal the disorder and why there is much concern in targeting these particular cells [57-58].

## 2.3 Clinical Presentation, Diagnosis and Staging

The presenting signs of pancreatic cancer depend on the position of the tumor within the gland, as well as on the stage of the disease. The bulk of tumors manifest in the head of the pancreas and produces obstructive cholestasis. Indefinite abdominal pain and nausea are also frequent. More infrequently, a pancreatic tumor may also produce duodenal difficulty or gastrointestinal bleeding. Pancreatic cancer often generates dull, deep upper abdominal discomfort that broadly restricts to the tumor region.

The difficulty of the pancreatic duct may commence to pancreatitis. Patients with pancreatic cancer often possess dysglycemia. Admittedly, pancreatic cancer should be judged in the differential diagnoses of severe pancreatitis and recently diagnosed diabetes.

At presentation, most victims also have systemic signs of the disorder such as asthenia, anorexia, and weight loss. Additional, less frequent signs combine intense and cosmetic venous thrombosis, panniculitis, liver-function abnormalities, gastric-outlet difficulty, increased abdominal girth, and depression.

A physical review may exhibit jaundice, temporal wasting, peripheral lymphadenopathy, hepatomegaly, and ascites. Outcomes of routine blood tests are frequently nonspecific and may involve clear irregularities in liver-function experiments, hyperglycemia, and anemia [23,47].

Evaluation of a victim in whom pancreatic cancer is presumed should concentrate on analysis and presentation of the disease, evaluation of resectability, and palliation of signs. Multiphase, multidetector helical computed tomography (CT) with intravenous management of diversity element is the imaging scheme of opportunity for the primary evaluation [59]. This procedure enables visualization of the fundamental tumor with the superior mesenteric artery, celiac axis, superior mesenteric vein, and portal vein and also with distant organs. In common, contrast-enhanced CT is adequate to strengthen a presumed pancreatic section and to construct an introductory supervision plan. Overall, contrast-enhanced CT prophesies surgical resectability with 80-90% efficiency [60]. Positron-emission tomography can be beneficial if the CT verdicts are ambiguous.

Several victims need supplementary diagnostic investigations. Endoscopic ultrasonography is beneficial in patients in whom pancreatic cancer is speculated although there is no obvious mass identifiable on CT. It is the favored technique of acquiring tissue for diagnostic objectives. Although a tissue diagnosis is not required in inmates who are programmed for surgery, it is expected before the introduction of treatment with chemotherapy or radiation therapy. Endoscopic retrograde cholangiopancreatography (ERCP) bestows the pancreatic and bile-duct anatomy and can be utilized to guide ductal brushing and lavage, which produces tissue for diagnosis. The ERCP technique is particularly beneficial in sufferers with jaundice in whom an endoscopic stent is needed to discharge obstruction [61]. In victims who possess large tumors, particularly in the body and tail of the pancreas, as well as other signs of advanced disease such as weight loss, a raised level of carbohydrate antigen 19-9 (CA 19-9), ascites, or equivocal CT findings, a staging laparoscopy can certainly discover metastatic and vascular reflection [62].

There are many possible plasma biomarkers for diagnosis, delamination of prognosis, and monitoring of therapy [63]. CA 19-9 is the only biomarker with confirmed clinical application and is beneficial for therapeutic monitoring and early exposure of recurrent infection after treatment in patients with known pancreatic cancer [63-67]. Nevertheless, CA 19-9 has significant shortcomings. It is not a particular biomarker for pancreatic cancer; the level of CA 19-9 may be raised in other conditions such as cholestasis. Besides, patients who are negative for Lewis antigen a or b (approximately 10% of patients with pancreatic cancer) are incapable to synthesize CA 19-9 and hold undetectable levels, even in advanced degrees of the disorder. Although the determination of serum CA 19-9 levels are beneficial in victims with known pancreatic cancer, the application of this biomarker as a screening tool has possessed inadequate outcomes.

Universal initial screening for pancreatic cancer is currently not supported, given the devices accessible and their execution [68]. Individual institution investigations concentrating on monitoring of victims at great danger, such as those with a powerful family history or cancer-predisposition syndromes, have adopted serial endoscopic ultrasonography and CT. Pancreatic lesions correlated with benign intrapancreatic mucinous neoplasia or pancreatic intraepithelial neoplasia have been identified in nearly 10% of these high-risk inmates. Nevertheless, the cost-effectiveness of this procedure is unclear, and its use is investigational [69].

TABLE 2.1: Staging of Pancreatic Cancer. (N denotes regional lymph nodes, M distant metastases and T primary tumor [71].

| Stage | Tumor Grade | Nodal Status | Distant Metastases | Medial Survival (months) | Characteristics |
|---|---|---|---|---|---|
| IA | T1 | N0 | M0 | 24.1 | Tumor limited to the pancreas, ≤2 cm in longest dimension |
| IB | T2 | N0 | M0 | 20.6 | Tumor limited to the pancreas, >2 cm in longest dimension |
| IIA | T3 | N0 | M0 | 15.4 | Tumor extends beyond the pancreas but does not involve the celiac axis or superior mesenteric artery |
| IIB | T1, T2 or T3 | N1 | M0 | 12.7 | Regional lymph-node metastasis |
| III | T4 | N0 or N1 | M0 | 10.6 | Tumor involves the celiac axis or the superior mesenteric artery (unresectable disease) |
| IV | T1, T2, T3 or T4 | N0 or N1 | M1 | 4.5 | Distant metastasis |

## 2.4 Staging of Pancreatic Cancer

Pancreatic cancer is staged according to the most up-to-date edition of the American Joint Committee on Cancer tumor-node-metastasis organization, which is based on the estimation of resectability utilizing helical CT [70]. T1, T2, and T3 tumors are conceivably resectable, whereas T4 tumors, which require the superior mesenteric artery or celiac axis, are unresectable (Table 2.1). Tumors affecting the superior mesenteric veins, portal veins, or splenic veins are categorized as T3 since these veins can be resected and reproduced, provided that they are patent.

## 2.5 Management of Early Disease

Victims with pancreatic cancer are best cared for by multidisciplinary teams that involve surgeons, medical and radiation oncologists, radiologists, gastroenterologists, nutritionists, and pain specialists, among others [72-73]. For victims with a resectable disorder, surgery remains the treatment of choice [74]. Depending on the position of the tumor, the operative methods may include cephalic pancreatoduodenectomy, distal pancreatectomy, or complete pancreatectomy. A least of 12 to 15 lymph nodes should be resected, and every endeavor should be made to achieve a tumor-free edge. Data from different randomized clinical tests designate that a more comprehensive resection does not increase survival but improves postoperative morbidity. Current investigations reveal that the consequences of vein resection and vascular reconstruction in patients with the inadequate engagement of the superior mesenteric vein and portal vein are alike to the outcomes in patients without vein engagement [75]. Poor prognostic factors involve lymph-node metastases, a high tumor grade, a large tumor, high levels of CA 19-9, persistently elevated postoperative levels of CA 19-9, and positive boundaries of resection [64,66,76-77].

Up to 70% of victims with pancreatic cancer present with biliary difficulty, which can be mitigated by percutaneous or endoscopic stent deployment. Decompression is suitable for patients in whom surgery is delayed, such as patients who are prescribed with neoadjuvant therapy before resection or who are assigned to other centers for treatment [78]. Victims with signs of cholangitis demand decompression as well as an antibiotic treatment before surgery.

Even if the tumor is completely resected, the consequence in patients with early pancreatic cancer is inadequate. The results of three large randomized clinical trials have confirmed the role of postoperative treatment in patients with resected pancreatic cancer [79-81]. The consequences of the European Study Group for Pancreatic Cancer Trial 1 and Charité Onkologie 1 trial confirm that postoperative administration of chemotherapy with either fluorouracil and leucovorin or gemcitabine, a nucleotide analogue generally employed to treat advanced pancreatic cancer, increases progression-free and overall survival. Also, the Radiation Therapy Oncology Group trial 97-04 revealed that the succession of gemcitabine with fluorouracil served as a continuous immersion and radiation therapy ended in a trend approaching expanded overall survival, although the improvement was not meaningful, among patients with tumors in the head of the pancreas. These outcomes are comparable to those of large single-institution series that combined radiation therapy [82].

In spite of diversity in patient communities and therapies, the consequence in patients treated in these trials was alike, with median durability of 20 to 22 months. Large tumor size, high differentiation grade, and preoccupation of the lymph nodes are risk factors for chronic disease. The consequence of positive resection boundaries, however, is more questionable [83]. Therefore, gemcitabine alone or gemcitabine in combination with fluorouracil based chemoradiation can be acknowledged the model of interest in this context. The straightforward explanation that postoperative treatment enhances the result in these victims is one of the most significant advancements that has been made in the management of pancreatic cancer.

An emerging approach in patients with resectable pancreatic cancer is the application of preoperative treatment. Nonrandomized, phase 2 investigations recommend that this method is at least as useful as postoperative treatment and may diminish the rate of local failures and positive resection boundaries after surgery [84]. These conclusions are expressly suitable for victims who have so-called borderline-resectable tumors with the restricted vascular association; in these cases, preoperative treatment may appear in tumor-free resection boundaries [85].

## 2.6 Incidence

The incidence of pancreatic cancer differs considerably over areas and communities. Incidence rates for pancreatic cancer in 2012 were highest in Northern America (7.4 per 100000 people) and Western Europe (7.3 per 100000 people), followed by other areas in Europe and Australia/New Zealand (uniformly about 6.5 per 100000 people) (Figure 2.2) [87]. The cheapest rates (about 1.0 per 100000 people) were mentioned in Middle Africa and South-Central Asia. Variations in incidence rates were twenty-fold among the communities with the highest rate (Czech Republic - 9.7), and the community with the lowest rate (Pakistan - 0.5). More than half of new cases (55.5%) were recorded in more advanced countries [87]. Somewhat less than half (41.0%; 139363 of events) of all current events of pancreatic cancer in 2012 were reported in the countries of Asia.

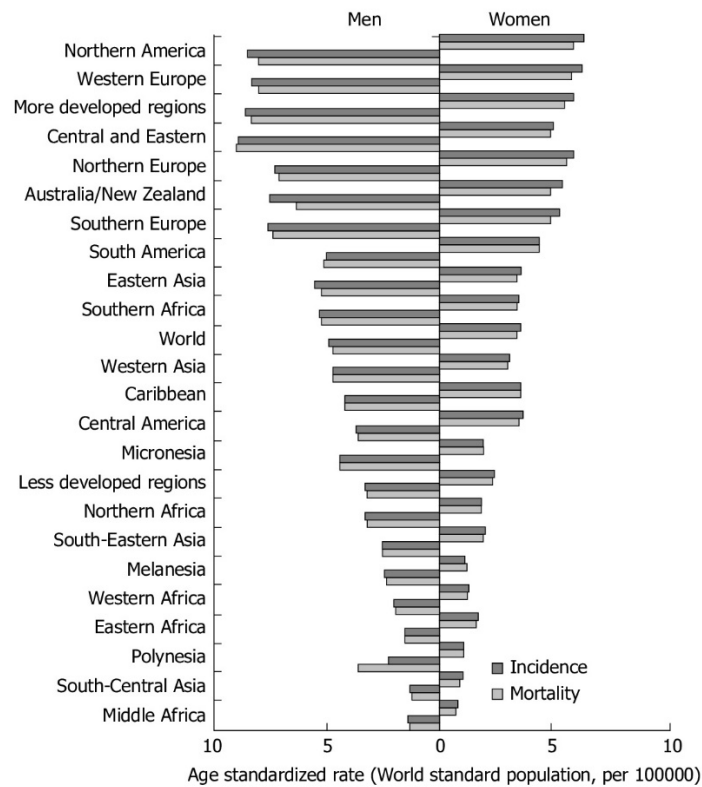

Figure 2.2: Pancreatic cancer incidence and mortality in men and women, by regions [86]

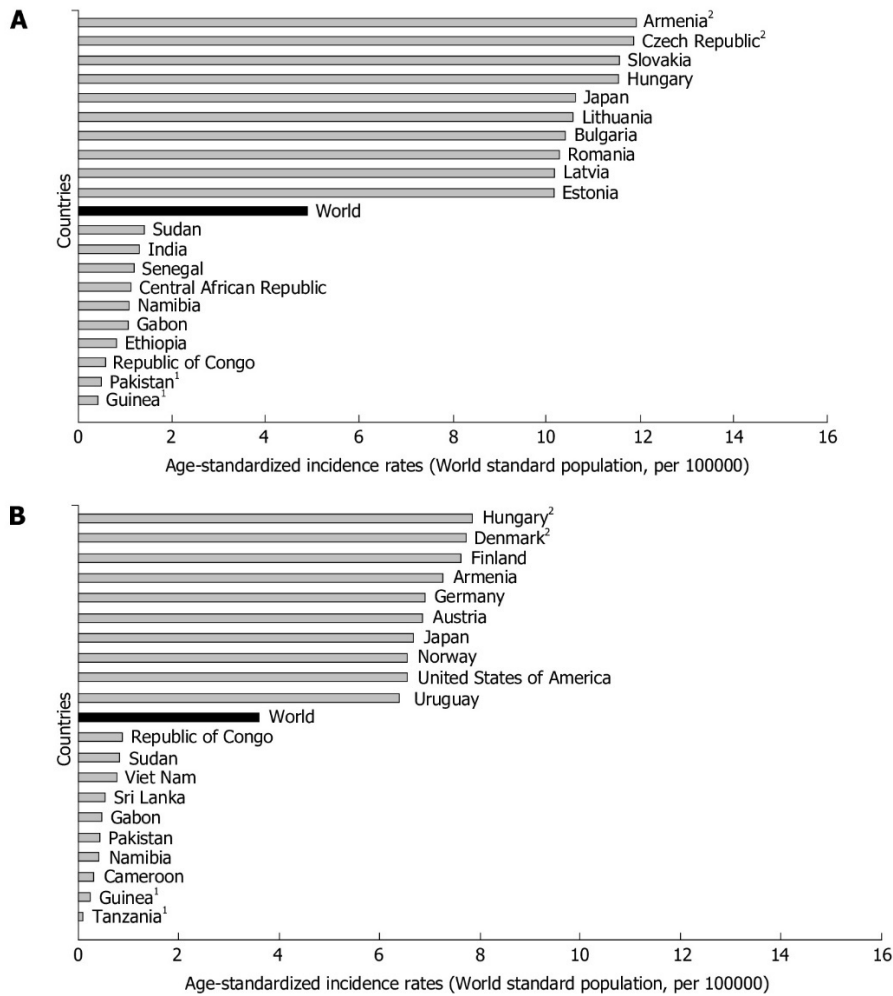

Figure 2.3: Pancreatic cancer incidence in men (A) and women (B) [86]

There are meaningful geographic differences in the incidence of pancreatic cancer by genders [87-88]. The incidence rate of pancreatic cancer among men in 2012 was 4.9 per 100000, and among women 3.6 per 100000 (Figure 2.3) [87]. In men, the chance of advancing pancreatic cancer was high in Armenia (11.9) and Czech Republic (11.8), Slovakia and Hungary (equally - 11.5), then in Japan and Lithuania (equally - 10.6) (Figure 2.3A). In contrast, the chance of incurring pancreatic cancer in men was the cheapest in Pakistan (0.5) and Guinea (0.4). The regions with the highest incidence rates of pancreatic cancer in women were Northern America (6.4 per 100000), Western Europe (6.3), and Northern Europe and Australia/New Zealand (5.9 and 5.4, respectively) (Figure 2.2) [87]. The lowest allowances of pancreatic cancer incidence (less than 1.0 per 100000) were in Middle Africa and Polynesia. Women living in Hungary, Denmark, Finland and Armenia have the most prominent risk (almost 7.5 per 100000) of dying from pancreatic cancer, while women in Tanzania, Guinea, Cameroon, Namibia and Pakistan have the lowest disease risk (less than 1.0 per 100000) (Figure 2.3B).

The incidence rates for both sexes develop with age, the highest in older than 70 years [89-90]. It is predominantly a disorder of the elderly, and almost 90% of all events are diagnosed after the age of 55 years [91].

Although it is not feasible to completely describe the variations in the incidence of pancreatic cancer in diverse parts of the world, most of the international variety in the incidence of pancreatic cancer has been associated to display to observed or assumed risk factors associated to lifestyle or the environment [92-93]. Tobacco smoking is likely to resolve these international distinctions and gender variations [94]. Some verdicts may be intimating that obesity may have some influence on differences [95]. Additionally, some conclusions show the purpose of aging and hereditary and genetic factors. The causes for a higher incidence of pancreatic cancer in men are still inadequately identified: women are either less prone to these kinds of destructive tumors, or are less exposed to risk factors from the circumstances responsible for their occurrence [92,96]. Besides that, international variations display diagnostic potential and the variation in the application of various diagnostic modalities [97]. In 2012, Europe stocked one-third of the overall incidence, which likely speculated the more specific diagnosis of pancreatic cancer rather than etiology [98]. It should be remarked that some variations in the incidence of pancreatic cancer around the world may be associated with the quality of registries, which coverage, completeness and correctness differs by country [99].

## 2.7 Mortality

International mortality rates for pancreatic cancer differ significantly in diverse places. Rates of mortality from pancreatic cancer in 2012 in both genders were highest in Northern America (6.9 per 100000 people) and Western Europe (6.8), supported by other European regions and Australia/New Zealand (almost 6.0, respectively) (Figure 2.2) [87]. The lowest mortality was reported in the countries in Middle Africa and South- Central

Asia (less than 1.0 per 1000000 people). The variations in mortality rates were nearly fifty-fold between the populations with the highest and lowest rate (Armenia vs Tanzania: 8.9 vs 0.2). More than one third (111029 deaths) of all departed from pancreatic cancer are citizens of Europe. Somewhat less than half (41.5%; 137251 deaths) of all deaths from pancreatic cancer were reported in 2012 in Asian countries [87]. More than half (55.8%, 184429 deaths) of expired of pancreatic cancer were recorded in more advanced regions. The least number of deaths was recorded in Micronesia/Polynesia.

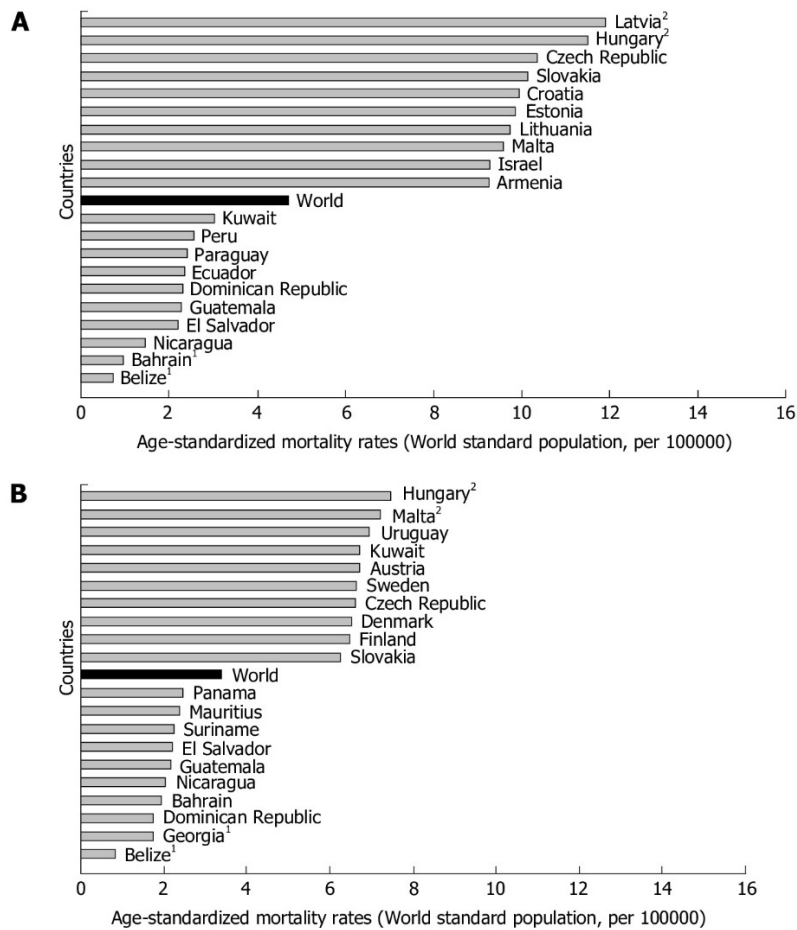

Figure 2.4: Pancreatic cancer mortality in men (A) and women (B) [86]

Mortality of pancreatic cancer in both genders progresses with age, and almost 90% of all deaths are recorded after the age of 55 years [87,89]. The highest death degrees in 2012 in males were reported in Central and Eastern Europe (Latvia - 11.9, Hungary - 11.5) (Figure 2.4A) [87]. The death from pancreatic cancer was lowest (less than 1.0 per 100000 people) in Belize and Bahrain. The highest death rates in 2012 in females were reported in Hungary (7.5) and Malta (7.2) (Figure 2.4B) [87]. The mortality from pancreatic cancer was lowest in women in Belize (0.8). Destruction of pancreatic cancer is nearly indistinguishable with its incidence because it is one of the deadliest malignant tumors [100-101]. Ideas for the substantial variations in death rates of pancreatic cancer were not entirely elucidated. Variations in rates of incidence can be apparent and apparent. Specious diversity may appear as a result of differences in the diagnosis of diseases and causes of death, as a result of a real transformation in the incidence and fatality. Data on the incidence/mortality issued by WHO are not of the same degree in all countries [99]. Although the quality (accuracy and completeness of reason of death registration, fundamentally) and the coverage of knowledge in most emerging countries can be estimated insufficient, the registry often settles the only accessible source. Symptoms, signs and inadequately specified requirements as the underlying cause of death are significantly more often introduced in Serbia, the Russian Federation and Greece than in more advanced countries such as the United States of America, United Kingdom, and Finland, which leads to the demand for a cautious interpretation of the data statistics of mortality in international comparisons [99]. Pancreatic cancer is tough to diagnose. The malignant pancreatic neoplasm was among the most common cancers identified at autopsy investigations [97,102]. It is comprehended that for pancreatic cancer there is no functional modality of screening, early detection and effective treatment, which has the consequence of survival rates diversifying very little between advanced and emerging countries [103]. Currently accessible treatment choices for pancreatic cancer are inadequate. Due to the advanced stage at diagnosis, 80%-90% of patients have unresectable tumors and long-term durability after surgical resection is reduced [100,104].

High smoking prevalence has been universally acknowledged as the main contributor to the high fatality rates of pancreatic cancer [94,105]. Diverse proof support that diet (animal fat and meat consumption, etc.) performs a function in the growth of pancreatic cancer [106-107]. Also, the highest rates for pancreatic cancer mortality in eastern European countries implied that other factors (including the prevalence of diabetes, obesity, alcohol consumption) could determine the mortality of pancreatic cancer [108].

## 2.8 Survival

Cancer of the pancreas remains one of the most lethal common cancer types: the Mortality/Incidence ratio is 98% [87]. The overall five-year survival rate is about 6% (ranges from 2% to 9%), but this slightly reveals varying data quality globally [109-110]. For pancreatic cancer, survival rates change very small between advanced and emerging countries [109].

Based on the United States National Cancer Institute data for pancreatic cancer in both sexes and all races, 9.4% are diagnosed at the local stage while the 5-year durability for the localized disease was 29.3% during 2006-2012 [3]. More than half (52%) of all cases were diagnosed at the sparse stage with the 5-year durability rates of 2.6%.

Several intercountry durability variations for pancreatic cancers exist over Europe: 5-year survival rate was less than 3% in both sexes in England and Wales [111], Denmark and Sweden - 3.8% [103], in Italy - 1.2% [112]. EUROCARE Working Group examined the durability of cancer victims diagnosed from 1990 to 1994 in 22 European countries and recorded that 5-year survival rates were highest in men in Estonia (7.0%) and women in the Czech Republic (7.5%), while lowest survival rates were documented in men in Malta (0.0%) and women in Slovenia (1.3%) [109]. Durability from pancreatic cancer in Germany in the early 21st century was 9.0% [86].

Survival rates of pancreatic cancer in community are influenced by many factors, such as the type of cancer, the staging at the time of diagnosis, serum albumin level, and tumor size, treatment modality, availability and diversity in health care systems, and other factors including age, sex, overall health, lifestyle [86]. Besides, pancreatic cancer survival rates could be controlled by factors such as the efficacy of cancer registry, exhaustiveness and nature of registration data, completeness of follow-up [86,103].

## 2.9 Conclusion

In this chapter, we have discussed clinical representation, diagnosis, staging, management of early disease, incidence, mortality and survival of pancreatic cancer.

CHAPTER 3
# Background Study and Literature Review

*Introduction*

*Deoxyribonucleic Acid (DNA)*

*Protein*

*Genes*

*Microarray Technology*

*MicroRNA Technology*

*Literature Review*

*Conclusion*

## 3.1 Introduction

This chapter presents the biological terms required for this research. DNA, protein, genes, microarray and microRNA technology have been introduced sequentially which explains the aspects of this research from a perspective of bioinformatics and statistics. Moreover, a literature review has been introduced with popular and recent researches and their findings. Finally, the domain of our research is discussed.

## 3.2 Deoxyribonucleic Acid (DNA)

Deoxyribonucleic Acid (DNA) is a molecule formed of two strings that twist around each other to form a dual helix bearing genetic guidance for the improvement, functioning, maturity and generation of all distinguished organisms and many viruses. DNA and ribonucleic acid (RNA) are nucleic acids; alongside proteins, lipids including complex carbohydrates (polysaccharides), nucleic acids are one of the four principal kinds of macromolecules that are necessary for all distinguished sorts of life.

DNA is a long polymer formed from recurring units called nucleotides [113-114]. The structure of DNA is dynamic along its length, being proficient of twisting into compact circles and other configurations [115]. In all classes, it is constituted of two helical strings, connected by hydrogen links. Both strings are coiled around the equivalent axis and possess the same pitch of 34 angstroms (Å) (3.4 nanometers). The pair of strings has a span of 10 angstroms (1.0 nanometer) [116]. According to another research, when surveyed in a separate clarification, the DNA chain estimated 22 to 26 angstroms wide (2.2 to 2.6 nanometers), and one nucleotide unit estimated 3.3 Å (0.33 nm) long [117]. Although each nucleotide is very inadequate, a DNA polymer can be very extensive and include hundreds of millions, such as in chromosome 1. Chromosome 1 is the most comprehensive human chromosome with nearly 220 million base pairs, [118] and would be 85 mm long if unfolded.

DNA does not normally endure as a particular strand, but alternately as a pair of strands that are compressed tightly together [116,119]. These two long strands twist around each other, in the appearance of a double helix. The nucleotide includes both a portion of the resolution of the molecule (that keeps the chain together) and a nucleobase (that combines with the other DNA strand in the helix). A nucleobase connected to sugar is termed a nucleoside, and a base connected to a sugar and one or more phosphate groups is termed a nucleotide. A biopolymer comprising multiple linked nucleotides (as in DNA) is named a polynucleotide [120].

The DNA double helix is maintained essentially by two forces: hydrogen bonds within nucleotides and base-stacking communications among aromatic nucleobases [121]. In the cytosol of the cell, the conjugated pi bonds of nucleotide bases regulate perpendicular to the axis of the DNA molecule, reducing their communication with the solvation shell. The four bases discovered in DNA are adenine (A), cytosine (C), guanine (G) and thymine (T). These four bases are associated with the sugar-phosphate to form the comprehensive nucleotide. Adenine couples with thymine and guanine partners with cytosine, producing A-T and G-C base pairs [122-123].

## 3.3   Protein

Proteins are vital nutrients for the human body [124]. They are one of the construction pieces of body tissue and can also work as a fuel origin. As a fuel, proteins render as much energy density as carbohydrates: 4 kcal (17 kJ) per gram; in opposition, lipids provide 9 kcal (37 kJ) per gram. The most significant character and a defining feature of protein from a nutritional viewpoint is its amino acid formation [125].

Proteins are polymer strings composed of amino acids connected by peptide bonds. Throughout human digestion, proteins are torn down in the stomach to smaller polypeptide chains through hydrochloric acid and protease reactions. This is significant for the absorption of the necessary amino acids that cannot be biosynthesized by the body [126].

There are nine fundamental amino acids which individuals must recover from their diet to restrict protein-energy consumption and resulting death. They are phenylalanine, valine, threonine, tryptophan, methionine, leucine, isoleucine, lysine, and histidine [125,127]. There has been a discussion as to whether there are 8 or 9 crucial amino acids [128]. The agreement appears to incline towards 9 considering histidine is not incorporated in adults [129]. There are five amino acids which humans can incorporate in the body. These five are alanine, aspartic acid, asparagine, glutamic acid and serine. There are six provisionally essential amino acids whose organization can be restricted under special pathophysiological situations, such as prematurity in the infant or individuals in critical catabolic disaster. These six are arginine, cysteine, glycine, glutamine, proline and tyrosine [125].

Dietary origins of protein combine both animals and plants: meats, dairy products, fish and eggs, as well as grains, legumes and nuts. Vegans can get adequate essential amino acids by consuming plant proteins [130].

## 3.4 Genes

In biology, a gene is a series of nucleotides in DNA or RNA that codes for a particle that has a purpose. While gene expression, the DNA is first transcribed into RNA. The RNA can be directly operative or be the intermediate template for a protein that plays a role. The transmission of genes to an organism's generation is the basis of the heritage of the phenotypic characteristic. These genes gain several DNA sequences termed genotypes. Genotypes along with environmental and developmental agents decide what the phenotypes will be. Most biological features are under the authority of polygenes as well as gene-environment communications. Some genetic attributes are instantly visible, such as eye color or a number of limbs, and some are not, such as blood type, the risk for particular disorders, or the thousands of fundamental biochemical manners that compose life.

Genes can obtain mutations in their progression, leading to several alternatives, known as alleles, in the population. These alleles encode somewhat altered versions of a protein, which induce altered phenotypical attributes. Convention of the term "having a gene" (e.g., "good genes," "hair color gene") typically refers to embracing a variant allele of the same, shared gene. Genes emerge due to natural selection/survival of the fittest and genetic disposition of the alleles.

The notion of a gene remains to be polished as innovative aspects are identified [131]. For example, administrative areas of a gene can be far separated from its coding domains, and coding areas can be divided into different exons. Some viruses stock their genome in RNA instead of DNA and some gene products are operative non-coding RNAs. Hence, a comprehensive, modern effective representation of a gene is any discrete locus of heritable, genomic distribution which influences an organism's habits by being exposed as an operative product or by the management of gene expression [132-133].

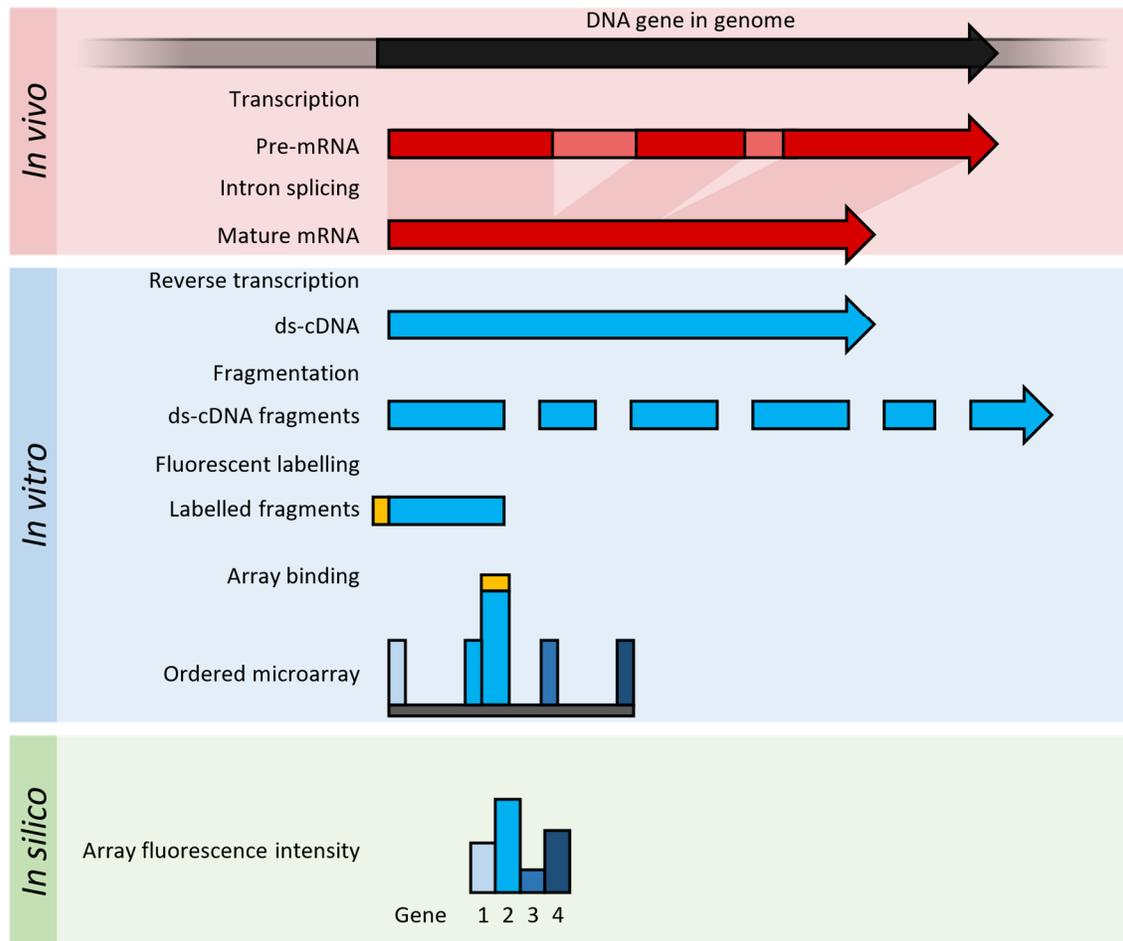

Figure 3.1: Summary of DNA Microarrays [134]

## 3.5 Microarray Technology

A DNA microarray (also generally recognized as DNA chip or biochip) is a combination of microscopic DNA points connected to a hard surface. Scientists utilize

DNA microarrays to estimate the expression levels of large quantities of genes concurrently or to genotype various regions of a genome. Each DNA spot holds picomoles (10-12 moles) of a particular DNA sequence, identified as probes. These can be a short segment of a gene or other DNA component that are applied to hybridize a cDNA or cRNA target following high stringency requirements. Probe-target hybridization is normally identified and quantified by detection of fluorophore, silver or chemiluminescence-labelled targets to decide relevant excess of nucleic acid chains in the target. The primary nucleic acid arrays were macro arrays generally 9 cm × 12 cm and the first automated image-based analysis was published in 1981 [135]. It was developed by Patrick O. Brown.

Figure 3.1 displays the summary of DNA Microarrays. Within the organisms, genes are reproduced and joined to provide mature mRNA transcripts (red). The mRNA is obtained from the organism and reverse transcriptase is utilized to copy the mRNA into permanent ds-cDNA (blue). In microarrays, the ds-cDNA is fragmented and fluorescently specified (orange). The labelled parts connect to an organized array of corresponding oligonucleotides, and determination of fluorescent intensity across the array shows the abundance of a proposed set of sequences. These sequences are typically particularly determined to arrive on genes of concern within the organism's genome [134].

The core law behind microarrays is hybridization between two DNA strands, the characteristic of corresponding nucleic acid sequences to correctly pair with each other by developing hydrogen linkages between corresponding nucleotide base pairs. A high number of interconnected base pairs in a nucleotide sequence indicates more established non-covalent bonding between the two strands. After cleaning off non-specific bonding sequences, only strongly paired strands will remain hybridized. Furthermore, labelled target sequences that connect to a probe sequence produce a signal that depends on the hybridization states (such as temperature) and cleaning after hybridization. The total power of the signal, from a point (feature), depends upon the number of target sample coupling to the probes present on that point. Microarrays use comparative quantitation in which the strength of a feature is compared to the strength of the same feature under a separate condition, and the identification of the feature is distinguished by its location.

## 3.6 MicroRNA Technology

MicroRNAs (miRNAs) are a group of inadequate (19-27 nt) noncoding RNAs proficient of base-paring to the reproductions of protein-coding genes (that are termed the targets of the miRNAs), pointing to downregulation or suppression of the targeted genes. The miRNA gene species is one of the most comprehensive in more distinguished eukaryotes: more than 700 adult miRNAs have been identified in the human genome, according to the current release of miRBase, and these miRNAs estimate for >2.5% of all human genes. The precise tools by which miRNAs coordinate their target genes expression remain unknown, although various models have been recommended, implying that miRNAs could influence translation control at both the preliminaries phase and the elongation phase of interpretation. Alternatively, the translation may not be directly concerned, but miRNAs could supervise to accelerated proteolysis of nascent polypeptides; or they may direct to the gathering of target mRNAs in the P-bodies, separating them from the translation machinations. The miRNAs could also point to the degradation of the target reproductions. Up-to-date testimony implies that miRNA-induced target transcript degeneration through a complicated manner that involves the deadenylation and decapping of mRNAs, distinguished from the siRNA-induced RNA silencing mechanism.

Computational prophecy of miRNA targets is extremely challenging in animals than in plants, because animal miRNAs frequently form defective base-pairing with their target sites, unlike plant miRNAs which nearly always treat their targets with near-perfect complementarity. In the past several ages, a large number of target prediction programs have been improved for animal miRNAs. Some of the most primitive target prediction programs (including TargetScan/TargetScanS, PicTar, miRanda, DIANA-microT and MicroInspector) utilized hand-derived habits based on a number of principles abstracted from distinguished miRNA-target interactions. These principles indicate: (a) near-perfect complementarity in the 6-8 nt region adjacent to the 50 end of the miRNA (the allegedly 'seed' zone) with the 30UTR section of the target sequence; (b) evolutionary maintenance of the target sequences between species; (c) effective thermodynamic balance of miRNA–mRNA duplex; (d) cooperativity between multiple sections in nearby proximity; and (e) presence of a central nonmatched zone (creating a loop). RNAhybrid, one of the earliest refined miRNA target prediction programs, performed an effective algorithm that obtains actively most favorable hybridizations between the miRNA and mRNA fragments, bypassing intramolecular hybridizations. A newer miRNA target prediction program, Rna22, implemented a method that first attains significant sequence motifs amongst all recognized miRNA sequences, then determines 'target islands', or sections in mRNAs where reverse complements of the miRNA patterns aggregate and converges on these target islands when hunting for miRNA target sites. Most recently exhibited miRNA target prediction programs (miTarget, MirTarget2 and NBmiRTar) applied machine learning techniques to assemble predictors directly from authorized miRNA target datasets. Besides, numerous up-to-date studies implied that target site receptiveness is an influential factor for efficient targeting of miRNAs. One program, PITA, was produced to perform predictions based on target site approachability characteristics [136].

## 3.7 Literature Review

As the survival rate of pancreatic cancer victims worldwide is the least for cancer-related survivals, therefore, researchers have been working on contributing in diverse stages of pancreas carcinoma, for example, finding biomarkers, proposing approaches and treatments for detection, prognosis, diagnosis and management of locally advanced disease. We studied several pieces of research that specifically focused on the analysis of pancreatic cancer to understand the problem domain and after that, we carefully picked our area of research.

A research of 2005 proposed that a meta-analysis is an influential tool for the identification and validation of differentially expressed genes (DEGs) and these can express good nominees for novel diagnostic and therapy targets for pancreatic cancer [37]. They applied meta-analysis for identification and RT-PCR for validation of DEGs. In 2006, clinically beneficial biomarkers were identified among mass spectrometry and quantitative proteomics by following the SILAC procedure as a screening tool for pancreatic cancer [137]. Several proteins not reported before to be praised in pancreatic cancer were also distinguished including CD9, perlecan, SDF4, apoE and fibronectin receptor. A study of 2013 revealed that impairment in a lipolytic pathway comprising lipases and a unique set of Free Fatty Acids (FFA) may perform an essential function in the advancement and improvement of pancreatic cancer and produce potential targets for the therapeutic intervention after conveying global metabolite profiling analysis on two independent cohorts of resected PDAC cases to distinguish critical metabolite alteration [138].

Meanwhile, researchers were attempting to contribute to prognosis as the diagnosis of pancreatic cancer was not exposing any improvement for survival. In 2014, a study reported that not only high miR-21 expression but also miR-23a and miR-27a have prognostic value for microRNA analysis and these individual miRNAs and their doublet expressions are, not surprisingly, highly correlated with the triple combination [139]. Merged gene expression analysis of whole-tissue and micro dissected pancreatic ductal adenocarcinoma was directed in 2016 to develop on other microarray studies of PDAC by putting together the higher statistical power due to a larger number of samples with knowledge about cell-type-specific expression and patient survival [140]. In 2017, various works displayed that carcinoembryonic antigen (CEA) along with carbohydrate antigen 19-9 (CA 19- 9) can predict the consequence of different diseases which includes pancreatic cancer [141] but certain biomarkers are insufficient to predict the outcome as it lacks proper sensitivity and specificity [142-143].

During the same time, a research suggested that when carbohydrate antigen 19-9 (CA 19-9) is combined with carcinoembryonic antigen (CEA) and carbohydrate antigen 125 (CA 125), CA 19-9 can assist predict the outcome of patients to surgery and chemotherapy [8]. Another study of pancreatic cancer determined molecular biomarkers using mRMR shortest path method for pancreatic cancer [5]. In 2018, finally, a study conducted on combined analysis of mRNA microarray and microRNA expression datasets suggested that statistical online tools can be employed for identifying prognostic biomarkers for pancreatic cancer [39]. However, statistical feature selection methods (parametric and non-parametric) may work better for identifying prognostic biomarkers. Besides, since there is still no specific diagnosis technique for pancreatic cancer, the survival rate has no advancement in decades. On the other hand, early diagnosis is the only way for survival improvement as previous research suggested [39], therefore, we selected identification of prognostic biomarkers as our domain of research.

## 3.8 Conclusion

In this chapter, we have reviewed DNA and protein and why they are essential for this study. The study of genes, microarray and microRNA technology paved the direction of the investigation. Literature review segment presented diverse studies and up-to-date researches for understanding the problem domain and deciding the domain of interest for our research.

# CHAPTER 4
# Materials and Dataset Processing

*Introduction*

*Gene Expression Omnibus*

*Research Datasets*

*P-Value*

*Upregulation and Downregulation*

*Fold-Change*

*Log-Transformation*

*Conclusion*

## 4.1 Introduction

This chapter first introduces Gene Expression Omnibus (GEO) [144], NCBI gene expression and hybridization array data repository. After that, the datasets used for this research are introduced. Moreover, P-value, upregulation and downregulation, fold-change and log-transformation are discussed as well.

## 4.2 Gene Expression Omnibus

The Gene Expression Omnibus (GEO) scheme was launched in acknowledgement to the increasing requirement for a public receptacle for high-throughput gene expression data. GEO presents a flexible and accessible design that promotes submission, storage and retrieval of heterogeneous data sets from high-throughput gene expression and genomic hybridization analyses.

GEO separates data into three principal components, platform, sample and series, each of which is accessioned (i.e. given a unique and permanent identifier) in a relational database. To obtain an extensive and adaptable design that enables storage and retrieval of very distinct data types, the data are not fully granulated within the database. Rather, a tab-delimited ASCII table is saved for each platform and each sample. The table consists of various columns with supplementing column header names. The data within this table are currently somewhat extracted for indexing but maybe moreover extracted for more comprehensive exploration and retrieval. Besides, any number of additional columns may be rendered by the submitter for the formation of additional, submitter-defined data.

An instance of a platform is, basically, a list of explorations that describe what set of molecules may be identified in any investigation employing that platform. For example, the platform data table may include GEO-defined columns recognizing the location and organic reagent contents of each probe such as a GenBank accession number, open reading frame (ORF) title and clone identifier, as well as submitter-defined information. Platform accession numbers have a 'GPL' prefix.

An instance of a sample represents the derivation of the set of molecules that are being investigated and employ platforms to produce atomic abundance data. Each sample has one, and only one, parent platform which must be beforehand specified. For example, a sample data table may include columns designating the ultimate, important affluence value of the corresponding point outlined in its platform, as well as any additional GEO-defined and submitter-defined information. Sample accession numbers have a 'GSM' prefix.

An instance of a series projects samples into the significant datasets which construct an investigation, and are connected by a general characteristic. Series accession numbers have a 'GSE' prefix.

## 4.3 Research Datasets

For this research, we have downloaded three mRNA microarray datasets: GSE15471 [145] (Whole-Tissue Gene Expression Study of Pancreatic Ductal Adenocarcinoma), GSE16515 [140] (Expression data from Mayo Clinic Pancreatic Tumor and Normal samples), GSE28735 [138] (Microarray gene-expression profiles of 45 matching pairs of pancreatic tumor and adjacent non-tumor tissues from 45 patients with pancreatic ductal adenocarcinoma) and one microRNA expression dataset: GSE41372 [139] (MicroRNAs Cooperatively Inhibit a Network of Tumor Suppressor Genes to Promote Pancreatic Tumor Growth and Progression). Table 4.1 displays a summary of all three mRNA microarray datasets and Table 4.2 displays summary of microRNA expression dataset.

### 4.3.1 GSE15471

Combinations of normal and tumor tissue samples were collected at the time of surgery from the resected pancreas of 36 pancreatic cancer victims. Gene expression was investigated on Affymetrix U133 plus 2.0 whole-genome microarrays. For three of the 36 normal-tumor sample couples, replicate microarray hybridizations were shipped out in order to estimate the technical estimation of fallacies. Thus, 78 GeneChip hybridizations were performed in total. A victim sample pair was suspended from further study as one of the individuals did not match the quality checks. The microarray data were consequently normalized by applying the Robust Multi-Array Average (RMA) algorithm. All data were submitted by AI and Bioinformatics Lab, Research department, ICI, Bucharest, Romania. The platform of the dataset used in this research is GPL570.

TABLE 4.1: mRNA Microarray Datasets Configuration at a glance.

| Datasets | Normal Samples | Tumor Samples | Total Samples | Number of Genes |
|---|---|---|---|---|
| GSE15471 | 36 | 36 | 72 | 54775 |
| GSE16515 | 16 | 36 | 52 | 54613 |
| GSE28735 | 45 | 45 | 90 | 28869 |
| Total | 97 | 117 | 214 | |

TABLE 4.2: microRNA Expression Dataset Configuration at a glance.

| Datasets | Normal Samples | Tumor Samples | Number of microRNAs |
|---|---|---|---|
| GSE41372 | 9 | 9 | 734 |

### 4.3.2 GSE16515

This experiment consists of 36 tumor samples and 16 normal samples; a total of 52 samples. 16 samples consist of both tumor and normal expression data, whereas 20 samples consist of only tumor data. All data were offered by Molecular Pharmacology and Experimental Therapeutics department, Mayo Clinic, Rochester, USA. The platform of the dataset applied in this research is GPL570.

### 4.3.3 GSE16515

90 total samples were investigated. Gene expression profile of 45 pairs of pancreatic tumor and neighboring non-tumor tissues were examined using Affymetrix GeneChip Human Gene 1.0 ST arrays. Tumor gene expression profiles were distinctly separate from non-tumor portraits. All data were presented by NCI/NIH, Bethesda, USA. The platform of the dataset adopted in this research is GPL6244.

### 4.3.4 GSE41372

Combinatorial interpretation of miRNA and mRNA expression in pancreatic ductal adenocarcinoma (PDAC)_mRNA. The microRNA dataset consists of pancreatic ductal adenocarcinoma tumor specimens of 9 victims along with normal pancreas specimens of 9 sound individuals. All data were presented by the Sapienza University of Rome, Rome, Italy. The platform of the dataset applied in this research is GPL16142, NanoString nCounter Human miRNA assay (v1).

## 4.4 P-Value

The soundness of the scientific outcome of a research paper should be based on more than the statistical investigation itself. Not only properly implemented statistical programs, but also an accurate representation of the statistical outcomes also performs a significant role in making the outcomes sound. To verify the importance of the study's outcome, the idea of statistical significance, typically imposed with an indication referred to as P-value is ordinarily practiced. The widespread adoption of P-values to summarize the outcomes of research studies could emerge from the improved size and complexity of data in current scientific research. A simple review of the research conclusions was required by both contributors and readers, which made the use of P-value more popular.

Since the initiation of P-value in 1900 by Pearson [146], the P-values are the favored approach to summarize the outcomes of medical studies. Since the P-value is the result of a statistical experiment, many scholars and readers acknowledge it as the most influential summary of the statistical investigations.

The P-value indicates the probability, for a presented statistical model that, when the null hypothesis is true, the statistical summary would be identical to or more utmost than the original perceived outcomes [147]. Hence, P-value only symbolizes how inconsistent the data are with a particular statistical model (usually with a null-hypothesis). The smaller the P-value, the more comprehensive statistical variance of the data with the null hypothesis. What is significant is that P values do not concentrate on the investigation hypothesis but the null hypothesis.

## 4.5 Upregulation and Downregulation

DNA damage seems to be the fundamental underlying cause of cancer [148-149]. If specific DNA restoration is lacking, DNA damages tend to expand. Unrepaired DNA damage can enhance mutational faults while DNA replication due to the error-prone translesion organization. DNA damage can also enhance epigenetic modifications due to failures while DNA restoration [150-151]. Such modifications and epigenetic changes can cause emergence to cancer. Thus, epigenetic downregulation or upregulation of DNA repairing genes is likely central to improvement to cancer [152-153].

As defined in Management of transcription in cancer, epigenetic downregulation of the DNA repairing gene MGMT occurs in 93% of bladder cancers, 88% of stomach cancers, 74% of thyroid cancers, 40%-90% of colorectal cancers and 50% of brain cancers [154]. Furthermore, epigenetic downregulation of LIG4 happens in 82% of colorectal cancers and epigenetic downregulation of NEIL1 happens in 62% of head and neck cancers and in 42% of non-small-cell lung carcinomas.

Epigenetic upregulation of the DNA repair genes PARP1 and FEN1 happen in numerous cancers. PARP1 and FEN1 are necessary genes in the error-prone and mutagenic DNA restoration pathway microhomology-mediated edge blending. If this pathway is upregulated, the excess modifications it produces can guide to cancer. PARP1 is over-expressed in tyrosine kinase-activated leukemias, in neuroblastoma, in testicular and other germ cell tumors, and in Ewing's sarcoma [154]. FEN1 is upregulated in the majority of cancers of the breast, prostate, stomach, neuroblastomas, pancreas, and lung [154].

## 4.6 Fold-Change

Fold change is a measure expressing how much a quantity varies fitting from an introductory to an ultimate value. For example, an opening value of 30 and a terminal value of 60 resembles to a fold change of 2 (or equivalently, a switch to 2 times), or in general words, a one-fold advance. Fold change is measured commonly as the proportion of the distinction between the ultimate value and the primary value over the fundamental value. Hence, if the primary value is A and the ultimate value is B, the fold change is (B - A)/A or equivalently B/A - 1. As another example, a shift from 80 to 20 would be a fold change of -0.75, while a change from 20 to 80 would be a fold change of 3 (a variation of 3 to 4 times the original).

Fold change is often practiced in the interpretation of gene expression data in microarray and RNA-Seq analyses, for estimating the variation in the expression level of a gene. A downside to and severe jeopardy of applying fold change in this environment is that it is biased and may avoid deferentially expressed genes with huge variations (B-A) but small ratios (A/B), pointing to a huge miss rate at large concentration.

Let's assume there are 50 read counts in control and 100 read counts in therapy for gene A. This means gene A is revealing twice in therapy or fold change is 2. This serves fine for overexpressed genes as the number directly resembles how many times a gene is overexpressed. But when it is another way around (i.e., treatment 50, control 100), the value of fold change will be 0.5 (all under-expressed genes will produce values between 0 to 1, while overexpressed genes will produce values from 1 to infinity). To make this surfaced, we apply log2 for representing the fold change i.e., log2 of 2 is 1 and log2 of 0.5 is -1.

## 4.7 Log-Transformation

The normal distribution is extensively employed in the fundamental and clinical investigation to model continuous results. Perversely, the symmetric bell-shaped arrangement usually does not sufficiently represent the examined data from investigation designs. Pretty frequently data appearing in real studies are so skewed that conventional analytical summaries of these data produce unreasonable conclusions. Many techniques have been disclosed to test the normality hypothesis of the examined data. When the distribution of the continuous data is non-normal, transformations of data are employed to secure the data as "normal" as possible and, therefore, extend the soundness of the associated statistical investigations. The log transformation is, arguably, the most widespread among the diverse kinds of transformations used to convert skewed data to relatively corresponding to normality.

If the original data supports a log-normal distribution or almost so, then the log-transformed data supports a normal or near-normal distribution. In this instance, the log-transformation does exclude or diminish skewness. Perversely, data resulting from many investigations do not approximate the log-normal distribution so implementing this transformation does not lessen the skewness of the distribution. In fact, in some instances employing the transformation can make the distribution more skewed than the fundamental data.

## 4.8 Conclusion

In this chapter, we have discussed the materials and processing methods needed for this research. First, the datasets have been introduced with proper definition and understanding. P-value and Fold-Change have also been discussed. Moreover, upregulation and downregulation along with log-transformation as a pre-processing method have been introduced in this chapter as well.

# CHAPTER 5
# Methodology

*Introduction*

*Student's T-Test*

*Kruskal-Wallis Test*

*Wilcoxon-Mann-Whitney Test*

*P-Value Adjustment*

*Several P-adjustment Approaches*

*miRecords*

*OncoLnc*

*Conclusion*

## 5.1 Introduction

To distinguish the Differentially Expressed Genes (DEGs) and Differentially Expressed MicroRNAs (DEMs), feature selection approaches are required since statistical feature selection procedures are suitable for gene expression data [155]. For this research, we have applied three feature selection approaches: Student's T-Test, Kruskal-Wallis Test and Wilcoxon-Mann-Whitney Test. This chapter discusses the mentioned approaches, P-value adjustment, several P-adjustment techniques, miRecords and OncoLnc.

## 5.2 Student's T-Test

Student's t-test is a parametric test based on the t-distribution. Student's distribution is titled in recognition of William Sealy Gosset (1876-1937), who initially ascertained it in 1908. He was one of the most authentic talents in contemporary science [156] and one of the best Oxford grads in chemistry and mathematics in his era.

During 1899, he considered a position as a brewer at Arthur Guinness Son & Co, Ltd in Dublin, Ireland. Serving for the Guinness brewery, he was involved in quality check based on small representations in several steps of the production manner. As Guinness restrained its agents from distributing any documents to restrict revelation of classified knowledge, Gosset had to issue his research under the pseudonym "Student" [157], and his correspondence was not recognized for remarkable period following the publication of his most notable accomplishments, therefore the distribution was titled Student's or t-distribution, leaving his contribution less well known than his outstanding outcomes in statistics.

His, presently, famous article "The Probable Error of a Mean" where he presented the t-test (originally, he described it as the z-test) [158], was initially disregarded by most statisticians for more than twenty years as the statistical society was not interested in small samples [159]. It was particularly R. Fisher who acknowledged the significance of Gosset's small-sample research, and who reconfigured and stretched it to two independent samples, correlation and regression, and presented the accurate estimation of degrees of freedom [160].

## 5.2.1 Two Independent Samples T-Test

While we distinguish the parameters of two groups (means, variances, or proportions), we need to differentiate two circumstances: samples may be dependent or independent according to how they were elected. Two random samples are independent if the sample picked from one group is not associated in any fashion to the sample from the other group. Assume, two random samples $X_{11}, X_{12}, \ldots, X_{1n_1}$ and $X_{21}, X_{22}, \ldots, X_{2n_2}$ where both of them reflects a normal distribution having means $\mu_1$ and $\mu_2$ along with variances $\sigma_1^2$ and $\sigma_2^2$ respectively. Applying t-test, hypothesis $H_0 : \mu_1 = \mu_2$ vs $H_0 : \mu_1 \neq \mu_2$ can be examined.

For equal variances ($\sigma_1^2 = \sigma_2^2 = \sigma^2$), it is a more simplistic condition since variances of examined groups, though unknown, are identical. With the corresponding sample sizes remaining $n_1$ and $n_2$, the maximum likelihood principle produces an analysis based on the test statistic

$$T = \frac{(\overline{X_1} - \overline{X_2}) - (\mu_1 - \mu_2)_0}{S_p \sqrt{\frac{1}{n_1} + \frac{1}{n_2}}} \tag{5.1}$$

where $S_p$ is the blended estimator of common variance $\sigma^2$ provided by,

$$S_p^2 = \frac{(n_1 - 1)S_1^2 + (n_2 - 1)S_2^2}{n_1 + n_2 - 2} \tag{5.2}$$

Here,

$$\overline{X_1} = \sum_{i=1}^{n_1} \frac{X_{1i}}{n_1}, \quad \overline{X_2} = \sum_{i=1}^{n_2} \frac{X_{2i}}{n_2},$$

$$S_1^2 = \frac{1}{n_1 - 1} \sum (X_{1i} - \overline{X_1})^2, \quad S_2^2 = \frac{1}{n_2 - 1} \sum (X_{2i} - \overline{X_2})^2$$

The combined t-test is based on the evidence that variable T in (5.1) has Student's relationships with $n_1 + n_2 - 2$ degrees of freedom, i.e., $P\left(t \geq t_{n_1+n_2-2}(\alpha)\right) = \alpha$. Therefore, for example, we decline the null hypothesis that both group means are identical ($H_0 : \mu_1 = \mu_2$) if $|t| \geq t_{n_1+n_2-2}(\alpha/2)$.

For unequal variances ($\sigma_1^2 \neq \sigma_2^2$), the test statistic is given by,

$$T = \frac{(\overline{X_1} - \overline{X_2}) - (\mu_1 - \mu_2)_0}{\sqrt{\frac{S_1^2}{n_1} + \frac{S_2^2}{n_2}}} \tag{5.3}$$

The distinction among the denominators in (5.1) and (5.3) should be regarded. In (5.1), we have the view of the common variance, while in (5.3) we have the appraisal of the variance of the distinction.

## 5.2.2 Robustness of T-Test

As the t-test expects certain assumptions in order to be accurate, it is of concern to understand how heavily the underlying hypotheses can be disrupted without diminishing the test outcomes considerably. In customary, a test is supposed to be robust if it is comparatively indifferent to the demolition of its underlying hypotheses. Therefore, a robust test is one in which the real value of significance is unchanged by failure to adhere hypotheses (i.e., it is near the trivial level of importance), and at the same time, the test keeps high strength.

Diverse researches have dealt with the sufficiency of the two-sample t-test if at least one hypothesis is disrupted. In the case of unequal variances, it has been confirmed that the t-test is particularly robust if sample sizes are equal [161]. Nevertheless, if two identical sample sizes are very small, the t-test may not be robust [162]. If both sample size and variances are uneven, the Welch t-test is favored to as a better scheme.

## 5.3 Kruskal-Wallis Test

When the dataset has outliers or doesn't support a normal distribution, a non-parametric test like Kruskal-Wallis Test [163] is adopted. In the statistical investigation, it is beneficial to apply ranks rather than the primary considerations, the principle on which Kruskal-Wallis test is developed. In the most elementary understanding, ranks simply organize all N measurements in order of quantity and replace the least by 1, second-least by 2 and so on. This provides the benefit of not possessing to consider a normal distribution of the measurements. For $C$ groups, where each group is of size $n_i$, the test statistic of Kruskal-Wallis is designated as

$$H = \frac{12}{N(N+1)} \sum_{i=1}^{C} \frac{R_i^2}{n_i} - 3(N+1) \tag{5.4}$$

Here,

$C$ is the number of groups,

$R_i$ is the sum of the ranks for the $i^{th}$ sample,

$n_i$ is the number of measurements of the $i^{th}$ sample,

$N = \sum n_i$ is the total number of measurements.

If there is no relation, that implies no two measurements are identical. If the value of H is extensive, it is refused by the null hypothesis. The hypothesis $H_0$ estimates that all C groups are from an equal community. However, most of the dataset appear from real life. Hence, a draw is very tangible. In the case of draws, each measurement is provided with the mean of the ranks for which it is drawn. Therefore, the test statistic H calculated from (5.4) is divided by

$$1 - \frac{\sum T}{N^3 - N}$$

For each group of draws, $T = (t-1)t(t+1) = t^3 - t$ and $t$ is the number of drawn measurements in the group. The summation is an overall common representation for the test statistic is obtained as

$$H = \frac{\frac{12}{N(N+1)} \sum_{i=1}^{C} \frac{R_i^2}{n_i} - 3(N+1)}{1 - \frac{\sum T}{N^3 - N}} \tag{5.5}$$

H for large samples is issued as $\chi^2(C-1)$, i.e. $\chi 2$ with $C-1$ degrees of freedom.

## 5.4 Wilcoxon-Mann-Whitney Test

The Wilcoxon-Mann-Whitney (WMW) test was introduced by Frank Wilcoxon in 1945 as "Wilcoxon rank sum test" and by Henry Mann and Donald Whitney in 1947 as "Mann-Whitney U test". Nevertheless, the test is more primitive: Gustav Deuchler introduced it in 1914 [164]. Now, this test is generally used as a nonparametric test for the two-sample situation dilemma. As with many other nonparametric tests, this is based on ranks rather than on the initial measurements.

The sample sizes of the two associations or random individuals are indicated by n and m. The measurements within each sample are independent and identically dispersed, and we assume independence between the two samples. The null hypothesis, $H_0$, is one of no distinction between the two organizations.

Let $F$ and $G$ be the distribution functions matching the two samples. Also, we have the null hypothesis $H_0: F(t) = G(t)$ for each $t$. Beneath the two-sided option, there is a distinction between $F$ and $G$. Often, it is believed that $F$ and $G$ are indistinguishable except a probable change in position, i.e., $F(t) = G(t-1)$ for every $t$. Then, the null hypothesis asserts $\theta = 0$, and the two-sided alternative is $H_0: \theta \neq 0$. Of course, one-sided options are feasible, too.

Let $V_i = 1$ while the $i^{th}$ smallest of the $N = n + m$ measurements are from the first sample and $V_i = 0$ oppositely. The Wilcoxon rank sum is a linear rank statistic expressed by $W = \sum_{i=1}^{N} i \cdot V_i$. Therefore, $W$ is the sum of the $n$ ranks of group 1; the ranks are decided based on the combined sample of all $N$ values. The Mann-Whitney statistic $U$ is described as $U = \sum_{i=1}^{n} \sum_{j=1}^{m} \Psi(X_i, Y_j)$ where $X_i(Y_j)$ is an inspection from group 1 (group 2), and

$$\Psi(X_i, Y_j) = \begin{cases} 1 & if \ X_i > Y_i \\ 0.5 & if \ X_i = Y_i \\ 0 & if \ X_i < Y_i \end{cases} \tag{5.6}$$

Because of $W = U + \frac{n}{2}(n+1)$, the experiments based on $W$ and $U$ are similar. The regulated statistic $Z_W$ can be calculated as $Z_W = \frac{W - E_0(W)}{\sqrt{Var_0(W)}}$ with $E_0(W) = \frac{n(N+1)}{2}$ and $Var_0(W) = \frac{nm(N+1)}{12}$. In the appearance of draws mean ranks can be confirmed for drawn observations. Then, the variance varies, in this case, we have

$$Var_0(W) = \frac{nm}{12}\left(N + 1 - \frac{\sum_{i=1}^{g}(t_i - 1)t_1(t_i + 1)}{N(N-1)}\right) \tag{5.7}$$

where $g$ is the number of drawn groups and $t_i$ the number of measurements within the $i^{th}$ drawn group. An untied value is considered as a drawn group with $t_i = 1$ [165].

Beneath $H_0$, the regulated Wilcoxon statistic asymptotically reflects a conventional normal distribution. This effect can be utilized to bring out the analysis and to measure an asymptotic P-value. According to Brunner and Munzel [166] the normal approach is satisfactory in case of $min(n, m) \geq 7$, if there were no draws. The two-sided asymptotic WMW test can discard $H_0$ if $|Z_W| \geq z_{1-\alpha/2}$, the corresponding P-value can be calculated as $2(1 - \Phi(|Z_W|))$, where $z_{1-\alpha/2}$ and $\Phi$ indicate the $(1 - \alpha/2)$-quantile and the partition function, sequentially, of the conventional normal distribution.

Alternatively, the specific alteration null distribution of W can be discovered and utilized for reasoning. Some monographs accommodate records of crucial values for the permutation test, but these tables can only be handled if there were no draws. A permutation test, nevertheless, is also reasonable in the presence of draws, because the specific qualified distribution of $W$ can be achieved.

As a rank test, the WMW test does not use all the accessible knowledge; instead of this, it is pretty influential. If the normal distribution is a sensible hypothesis, little is wasted by applying the Wilcoxon test rather than the parametric t-test. Again, when the hypothesis of uniformity is not settled, the nonparametric Wilcoxon test may have substantial improvements in times of efficiency. To be accurate, the asymptotic relative efficiency (ARE) of the WMW test in opposition to Student's t-test cannot be more petite than 0.864. Nonetheless, there is no upper boundary. If the data support a normal distribution then, ARE is $3/\Pi = 0.955$ [167].

The two-sided WMW test is compatible against all options with $\Pr(X_i < Y_j) \neq 0.5$. Nevertheless, the WMW test can provide a meaningful outcome for a test at the 5% level with considerably more than 5% possibility when the group medians are indistinguishable, but the group variances change. A generalization exists that can be utilized for testing a variation in position irrespective of a potential variation in variability [168].

## 5.5 P-Value Adjustment

If one requires to statistically understand whether a conclusion is meaningful, one may quantify the likelihood of that outcome transpiring by pure fortuitous possibility provided the null hypothesis. A traditional and spontaneous trade-off to discard the null hypothesis (therefore a significant non-random play) is 0.05 [169]. However, for our research, we have considered the adjusted P-value as 0.01 as suggested in previous researches [39]. Consequently, if the likelihood of examining the null hypothesis of uniformity of the mean of normalized expression levels of gene X and control groups ($\mu_1, \mu_2$) is less than 0.01, one would absolutely arbitrarily believe that it is their eureka moment by discarding the null hypothesis ($\mu_1 = \mu_2$), and encompassing the alternative hypothesis ($\mu_1 \neq \mu_2$). The most significant concern that remains in P-value quantification is during the occurrence of multiple measurements. For the statistical judgment of multiple associations, it may perform two main kinds of flaws that are indicated as Type I and Type II faults, respectively. The Type I fault is that one mistakenly discards a correct hypothesis, whereas Type II fault is referred to as a false negative. Since the specific amount of Type I and Type II faults are unobservable, one would expect to manage the likelihood of engaging these faults below a satisfactory level. In general, the managed expectations of performing Type I and Type II faults are negatively associated. Hence, one must find a relevant trade-off according to different experimental features and research objectives. If a meaningful outcome has significant functional importance, such as to maintain an efficient new operation, one would manage Type I fault more rigorously. However, one should bypass performing too many Type II faults when it intends to achieve fundamental nominees for distant investigation, which is very general in studies of genomics. Here, the managing of Type I fault is addressed as it considerably improves for multiple estimations.

In statistical conclusion, a probability value or P-value is directly or indirectly measured for every hypothesis and later matched with the pre-specified importance level $\alpha$ for deciding if this hypothesis should be discarded or not [170]. Hence, there are two ways of fixing the statistical conclusion of multiple measurements. First, it could directly regulate the perceived P-value for each hypothesis and retain the pre-specified importance level unchanged and this is hereby designated to as the adjusted P-value. Second, an adjusted cut-off corresponding to the originally pre-specified $\alpha$ could be also computationally measured and then matched with the witnessed P-value for statistical reasoning. In general, the adjusted P-value is more beneficial as here the tangible importance level is employed. For P-value adjustment, there exist several well-established methods.

## 5.6 Several P-adjustment Approaches

**Bonferroni Adjustment**

Bonferroni adjustment is one of the most regularly practiced methods for multiple measurements [171]. This approach attempts to manage the Family-Wise Error Rate (FWER) in a very powerful criterion and measure the adjusted P-values by directly multiplying the number of concurrently examined hypotheses ($m$):

$$p'_i = \min \{p_i \times m, 1\}(1 \leq i \leq m) \tag{5.8}$$

Bonferroni adjustment has been well-accepted to be much traditional particularly when there are a huge number of hypotheses being concurrently tested or hypotheses are extremely correlated.

**Holm Adjustment**

Based on Bonferroni adjustment, Holm adjustment was consequently introduced with less traditional persona [172]. Holm scheme, in a stepwise fashion, measures the importance levels depending on the P-value based rank of hypotheses. This stepwise analysis is beyond the extent of this study. For the $i^{th}$ ordered hypothesis $H_{(i)}$, the precisely adjusted importance level is computed:

$$\alpha' = \frac{\alpha}{m - i + 1} \tag{5.9}$$

The perceived P-value $P_{(i)}$ of hypothesis $H_{(i)}$ is later matched with its identical $\alpha'_{(i)}$ for statistical reasoning; and each hypothesis will be examined in order of the smallest to largest P-values $H_{(1)}, \ldots, H_{(m)}$. The association will instantly quit when the first $P_{(i)} \geq \alpha'_{(i)}$ is witnessed ($H = 1, \ldots, m$) and therefore all surviving hypotheses of $H_{(i)}$ ($j = i, \ldots, m$) are directly reported non-significant without demanding individual correspondence.

## Hochberg Adjustment

Hochberg P adjustment is a well-recognized method for P adjustment. To calculate the incorporated importance levels, it applies the formula asserted at Hochberg et al. [173]. The adjusted importance level for $j^{th}$ ordered hypothesis $H_{(j)}$ is estimated as

$$\alpha'_{(j)} = \frac{\alpha}{m-j+1} \tag{5.10}$$

This method distributes statistical conclusion of the hypothesis by starting with the largest p value $(H_{(m)}, \ldots, H_{(1)})$. When it is witnessed that $P_{(i)} < \alpha'_{(i)}$ for hypothesis $H_{(j)}$ ($j = m, \ldots, 1$), the comparison stops. Later we can assume that the hypothesis of $H_{(k)}$ ($k = j, \ldots, 1$) will be declined at significance level α.

## Hommel Adjustment

Simes [1986] modified Bonferroni approach and introduced a global analysis of m hypotheses [174]. Let $H = (H_{(1)}, \ldots, H_{(m)})$ be the global crossing hypothesis, H will be discarded if $p_{(i)} \leq i\alpha/m$ for any $i = 1, \ldots, m$. Nevertheless, Simes global test could not be utilized for evaluating the particular hypothesis $H(i)$). Hence, Hommel [1988] stretched Simes' method for examining individual $H(i)$ [175]. Let an index of $j = max\{i \in \{1, \ldots, m\} : P_{(m-i+k)} > k\alpha/i \text{ for } k = 1, \ldots, i\}$ be the size of the largest subset of $m$ hypotheses for which Simes test is not meaningful. All $H_{(i)}$ ($i = 1, \ldots, m$) are discarded if $j$ does not exist, otherwise, discard all $Hi$ with $P_i \leq \alpha/j$. Although a straightforward interpretation for measuring the adjusted P-values of Hommel method would not be simple, this job can be conveniently accomplished by computing tools, such as the p.adjust() function in R stats package.

## Benjamini-Hochberg (BH) Adjustment

In opposition to the powerful control of FWER, Benjamini and Hochberg [1995] proposed an arrangement for managing the false discovery rate (FDR), which is herein termed BH adjustment [176]. Let $q$ be the pre-specified upper bound of FDR (e.g., $q = 0.01$), the first step is to measure index $k$:

$$k = \max\{i: P_{(i)} \leq \frac{i}{m}q\} \tag{5.11}$$

If $k$ does not exist, deny no hypothesis, otherwise decline hypothesis of $H_{(i)}$ $(i = 1, ..., k)$. BH method begins with matching $H_{(i)}$ from the largest to smallest P-value $(i = m, ..., 1)$.

**Benjamini and Yekutieli (BY) Adjustment**

Alike BH approach, a more stable adjustment was further suggested for managing False Discovery Rate (FDR) by Benjamini and Yekutieli [2001], and this process is also termed as BY adjustment [177]. Let again $q$ be the pre-specified upper bound of FDR, the index $k$ is measured as:

$$k = \max\{i: P_{(i)} \leq \frac{i}{m}\tilde{q}\} \; with \; \tilde{q} = \frac{q}{\sum_{i=1}^{m}\frac{1}{i}} \tag{5.12}$$

If $k$ does not exist, deny no hypothesis, otherwise decline hypothesis of $H_{(i)}$ $(i = 1, ..., k)$. BY method can discuss the dependency of hypotheses with improved benefits.

TABLE 5.1: Comparison in data content between TarBase and miRecords [136]

|  | Total number of miRNAs | Total number of target genes | Total number of records | Number of low-throughput records | Number of human miRNAs | Number of human target genes | Number of human records |
|---|---|---|---|---|---|---|---|
| TarBase | 128 | 570 | 626 | 279 | 62 | 415 | 458 |
| miRecords | 301 | 902 | 1135 | 639 | 125 | 651 | 778 |

## 5.7 miRecords

MicroRNAs (miRNAs) are an essential group of small noncoding RNAs competent in controlling other genes expression. Much advancement has been performed in computational target prophecy of miRNAs in recent times. miRecords [136] is a resource for animal miRNA-target cooperation that consist of two parts. The Validated Targets component is a comprehensive, high-quality database of analytically verified miRNA targets emerging from precise standard research curation. As the largest observed accumulation of analytically verified miRNA targets, it highlights well-organized and structured documentation of empirical support of miRNA-target cooperation. This database not only assists the experimental researchers by presenting the lists of validated targets of the miRNAs of their affair but also presents a comprehensive and high-quality dataset that will expedite the improvement of the next-generation miRNA target prediction applications. The Predicted Targets component of miRecords is an alliance of prophesied miRNA targets generated by 11 authenticated miRNA target prediction programs. As the most comprehensive combination of prophesied miRNA targets, it is assumed to produce significant assist to researchers studying new miRNA targets.

TABLE 5.2: Comparison in miRNA target prediction programs executed among combined miRNA target resources, Part-A [136]

| Tool | DIANA-microT | Micro Inspector | mi-Randa | mi-Target | Mir-Target2 | NBmir-Tar |
|---|---|---|---|---|---|---|
| miRGator | | | ✓ | | | |
| miRGen | ✓ | | ✓ | | | |
| miRNAMap | | | ✓ | | | |
| miRecords | ✓ | ✓ | ✓ | ✓ | ✓ | ✓ |

TABLE 5.2: Comparison in miRNA target prediction programs executed among combined miRNA target resources, Part-B [136]

| Tool | Pic-Tar | PITA | RNA 22 | RNA Hybrid | TargetScan/TargetScanS |
|---|---|---|---|---|---|
| miRGator | ✓ | | | | ✓ |
| miRGen | ✓ | | | | ✓ |
| miRNAMap | | | | ✓ | ✓ |
| miRecords | ✓ | ✓ | ✓ | ✓ | ✓ |

TarBase is a famous database of analytically verified miRNA targets. Comparison between TarBase and miRecords (the Validated Targets component) recommends that miRecords hosts a much more comprehensive compilation of miRNA–target interactions than TarBase (Table 5.1). Furthermore, miRecords selects a compatible and systematic design of documenting the empirical guide for miRNA–target interactions, in which endogenous miRNA measures are separated from exogenous miRNA measures, three levels of empirical confirmation-the target gene level, the target region level and the target site level are clearly distinguished, and organizations of reporter assays, mRNA-level and protein-level computations and target-site-mutation knowledge are systematically documented. This is in contradiction to the documenting scheme embraced by TarBase, which only broadly incorporates all targets as either 'translationally repressed targets' or 'downregulated/cleaved targets', and contributes only limited knowledge about 'single site sufficiency', 'direct support' and 'indirect support' for the miRNA–target interactions.

The Predicted Targets component of miRecords combines the miRNA targets prophesied by 11 enduring miRNA target prediction programs, outnumbering all other combined reserves of predicted miRNA targets, including miRGator, miRGen and miRNAMap (Table 5.2, 5.3).

## 5.8 OncoLnc

OncoLnc [178] stocks over 400,000 inquiries, which combines Cox regression outcomes as well as mean and median representation of each gene. For the Cox regression outcomes, in accession to P-values, OncoLnc stocks the rank of the association. Various cancers bear very diverse P-value arrangements [178], and it is unclear what produces this distinction. As a result, employing one P-value cutoff across cancers is not feasible, and the rank of the association is a manageable way to estimate the comparative power of the correlation. The rank is determined per cancer, per data type.

The mRNA and miRNA identifiers employed by TCGA are out of date, and the identifiers in OncoLnc have been manually curated utilizing NCBI Gene: http://www.ncbi.nlm. nih.gov/gene, and recent miRBase outlines: http://www.mirbase.org with more than 2,000 updated mRNA representatives. Genes which have had their Entrez Gene ID eliminated from NCBI Gene, or could not be positively mapped to an individual identifier, are not incorporated in OncoLnc.

## 5.9 Conclusion

In this chapter, we have introduced several parametric and non-parametric feature selection approaches by applying which differentially expressed genes and differentially expressed microRNAs can be calculated. Furthermore, we have discussed P-value adjustment, several P-adjustment approaches, miRecords and OncoLnc.

# CHAPTER 6
# Implementation and Experimental Analysis

*Introduction*

*Experimental Setup*

*Workflow Diagram*

*Data Preprocessing*

*Implementation of Feature Selection Approaches*

*Prediction of microRNA Targets*

*Overall Survival Analysis*

*Comparative Study*

*Conclusion*

## 6.1 Introduction

This chapter describes the whole experimental procedure for our research. The chapter starts with the clarification of the experimental setup. After that, workflow diagram, data pre-processing, implementation of feature selection approaches, prediction of microRNA targets, overall survival analysis and comparative study are presented.

## 6.2 Experimental Setup

The feature selection methods have been developed in R programming language with RStudio, version 1.1.456. The other snippets for gene matching and further analysis were implemented with Python 3 in Anaconda, version 2018.12. The operating system of the computer was Windows 10. All the programs have been developed with the same personal computer (PC). The configuration of the PC is:

- ❖ Processor: Intel(R) CoreTM i3-7100U CPU @ 2.40GHz
- ❖ Installed Memory (RAM): 4.00 GB (3.89 GB usable)
- ❖ System Type: 64-bit Operating System, x64-based processor

All the resources, programs and snippets of our literature can be discovered at: https://github.com/Srizon143005/PancreaticCancerBiomarkers

## 6.3 Workflow Diagram

Figure 6.1 displays the workflow diagram of our experiment. We first pre-processed all four mRNA microarray and microRNA expression datasets. Then we applied feature selection methods on mRNA microarray datasets to identify differentially expressed genes (DEGs).

In the same process, we have selected differentially expressed microRNAs (DEMs) by applying feature selection methods on microRNA expression datasets. Then we identified the target genes for the DEMs and these target genes were matched with the DEGs. Finally, overall survival analysis was performed to identify the prognostic biomarkers of pancreatic cancer.

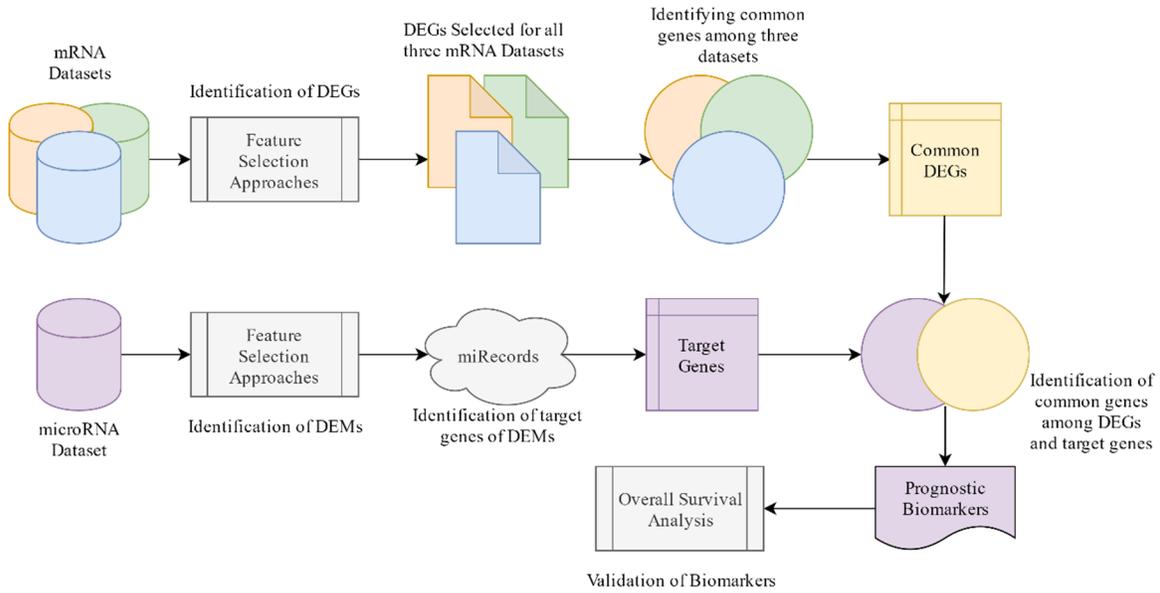

Figure 6.1: Workflow diagram of our research

## 6.4 Data Preprocessing

First, we downloaded three mRNA microarray datasets (GSE15471, GSE16515 and GSE28735) and one microRNA expression dataset (GSE41372) from Gene Expression Omnibus. GSE15471 and GSE16515 were based on GPL570 platform, GSE28735 was based on GPL6244 platform and GSE41372 was based on GPL16142 platform. All the datasets were downloaded in text format. These text datasets were then converted to CSV format (Comma Delimited) where the columns denote the samples and rows denotes genes expressed. We used GEO2R, an online tool to obtain log fold-change value for all datasets. While selecting log-fold change value the samples were identified as control and cancer groups and all other settings were left as default in GEO2R. Log-transformation was applied by GEO2R on GSE16515 and GSE41372. Our datasets didn't have any missing or null values; therefore, no further pre-processing was required.

## 6.5 Implementation of Feature Selection Approaches

After obtaining the values of log fold-change value, it's time to implement feature selection approaches on the datasets. Since GEO2R applied log-transformation on GSE16515 and GSE41372 while obtaining log fold-change values, we, therefore, applied log-transformation on these datasets before applying feature selection algorithms. Three feature selection algorithms: Student's T-Test, Kruskal-Wallis Test and Wilcoxon-Mann-Whitney Test were applied sequentially. For each test, P-value adjustment was performed using Holm, Hochberg, Hommel, Bonferroni, BH, BY p-adjustment approaches. We obtained the results for no p-adjustment as well. For applying the three feature selection methods we used the R functions t.test, kruskal.test and wilcox.test and for adjusting P-values p.adjust function was used. After obtaining the p-adjustment values for different methods, we selected only those genes (GSE15471, GSE16515 and GSE28735) and microRNAs (GSE41372) that had a p-adjustment value less than 0.01 and absolute log fold-change value higher than 1. These genes and microRNAs were identified as differentially expressed genes (DEGs) and differentially expressed microRNAs (DEMs).

TABLE 6.1: Number of identified of DEGs in mRNA datasets applying different methods and p-adjustment (Here, T, K and W refers to Student's T-Test, Kruskal-Wallis Test and Wilcoxon-Mann-Whitney Test respectively)

| Dataset | Method | Holm | Hochberg | Hommel | Bonferroni | BH | BY | None |
|---------|--------|------|----------|--------|------------|------|------|------|
| GSE15471 | T | 1630 | 1630 | 1706 | 1604 | 2595 | 2477 | 2617 |
|          | K | 1199 | 1199 | 1281 | 1179 | 2601 | 2380 | 2618 |
|          | W | 1396 | 1396 | 1485 | 1378 | 2604 | 2399 | 2622 |
| GSE16515 | T | 393  | 393  | 394  | 393  | 1830 | 1081 | 2439 |
|          | K | 43   | 43   | 51   | 43   | 2055 | 1034 | 2442 |
|          | W | 308  | 308  | 308  | 308  | 2048 | 1251 | 2444 |
| GSE28735 | T | 273  | 273  | 275  | 273  | 436  | 407  | 440  |
|          | K | 248  | 248  | 250  | 245  | 437  | 392  | 443  |
|          | W | 274  | 274  | 275  | 273  | 437  | 398  | 443  |

In Table 6.1, we have displayed the number of differentially expressed genes selected for corresponding feature selection algorithms for each p-adjustment method. Table 6.2 displays the number of upregulated, downregulated and total differentially expressed microRNAs selected for different feature selection algorithms for each p-adjustment method as well.

TABLE 6.2: Number of DEMs identified for micorRNA datasets applying different methods and p-adjustment (Here, up, down and WMW Test refers to upregulated DEMs, downregulated DEMs and Wilcoxon-Mann-Whitney Test respectively)

|  | Student's T-Test | | | Kruskal-Wallis Test | | | WMW Test | | |
| --- | --- | --- | --- | --- | --- | --- | --- | --- | --- |
|  | Up | Down | Total | Up | Down | Total | Up | Down | Total |
| Holm | 1 | 0 | 1 | 0 | 0 | 0 | 0 | 0 | 0 |
| Hochberg | 1 | 0 | 1 | 0 | 0 | 0 | 0 | 0 | 0 |
| Hommel | 1 | 0 | 1 | 0 | 0 | 0 | 0 | 0 | 0 |
| Bonferroni | 1 | 0 | 1 | 0 | 0 | 0 | 0 | 0 | 0 |
| BH | 14 | 2 | 16 | 0 | 0 | 0 | 0 | 0 | 0 |
| BY | 1 | 0 | 1 | 0 | 0 | 0 | 0 | 0 | 0 |
| None | 45 | 16 | 61 | 44 | 16 | 60 | 42 | 16 | 58 |

TABLE 6.3: The number of Upregulated and Downregulated DEGs for mRNA datasets

| Dataset | Upregulated DEGs | Downregulated DEGs | Total DEGs |
| --- | --- | --- | --- |
| GSE15471 | 2281 | 314 | 2595 |
| GSE16515 | 1480 | 350 | 1830 |
| GSE28735 | 271 | 165 | 436 |

If closely examined, it can be noticed that Kruskal-Wallis test identified the least number of DEGs for all three mRNA datasets, Wilcoxon-Mann-Whitney Test is in second place and Student's T-Test on third place and least number of DEGs was selected for Bonferroni correction. However, for identification of DEMs, no DEMs were selected by Kruskal-Wallis Test or Wilcoxon-Mann-Whitney Test for any p-adjustment. Therefore, we selected Student's T-Test for selection of DEGs and further studies. Moreover, it can be noticed that Bonferroni p-adjustment approach identified the least number of DEGs for all three mRNA microarray datasets. However, only one upregulated DEM was identified for Bonferroni p-adjustment approach. From previous research, we know that more than one microRNA is associated with the prognosis of pancreatic cancer which has been described in chapter 3 under the literature review section. Hence, we selected the second-best p-adjustment method, Benjamini-Hochberg p-adjustment for identifying DEGs and DEMs. This time 16 DEMs were identified.

TABLE 6.4: The identified DEMs, their p-values, adjusted p-values and log fold-change values

| Title of DEMs | P-values | Adjusted P-values | Log FC value |
|---|---|---|---|
| Upregulated DEMs | | | |
| hsa-miR-135b | 8.92E-07 | 0.000655 | 2.819987 |
| hsa-miR-197 | 3.25E-05 | 0.006422 | 2.192276 |
| hsa-miR-490-3p | 3.62E-05 | 0.006422 | 1.888625 |
| hsa-miR-221 | 4.77E-05 | 0.006422 | 2.181648 |
| hsa-miR-145 | 5.82E-05 | 0.006422 | 3.244704 |
| hsa-miR-10a | 6.12E-05 | 0.006422 | 2.841119 |
| hsa-miR-484 | 8.17E-05 | 0.007498 | 1.835517 |
| hsa-miR-337-3p | 0.000138 | 0.009322 | 1.307959 |
| hsa-miR-342-3p | 0.000147 | 0.009322 | 2.136132 |
| hsa-miR-223 | 0.000153 | 0.009322 | 2.930734 |
| hsa-miR-140-3p | 0.000165 | 0.009322 | 1.504247 |
| hsa-miR-21 | 0.00021 | 0.00971 | 2.992536 |
| hsa-miR-331-3p | 0.000212 | 0.00971 | 1.830589 |
| hsa-miR-190b | 0.000231 | 0.009977 | 1.369034 |
| Downregulated DEMs | | | |
| hsa-miR-302c | 4.31E-05 | 0.006422 | -1.37368 |
| hsa-miR-630 | 0.000201 | 0.00971 | -4.21846 |

From Table 6.3 it can be observed that 2595 (2281 upregulated and 314 downregulated in GSE15471), 1830 (1480 upregulated and 350 downregulated in GSE16515) and 436 (271 upregulated and 165 downregulated in GSE28735) DEGs were identified in pancreatic carcinoma tissue against normal tissue. After selecting DEGs separately in three mRNA datasets, we selected only those DEGs which were present in all three mRNA datasets. In particular, a total of 178 DEGs were selected. Among them, 152 were upregulated and 26 were downregulated. Venn Diagram was then plotted to show common DEGs in all three mRNA datasets for upregulated and downregulated genes separately. Figure 6.2 displays the illustration.

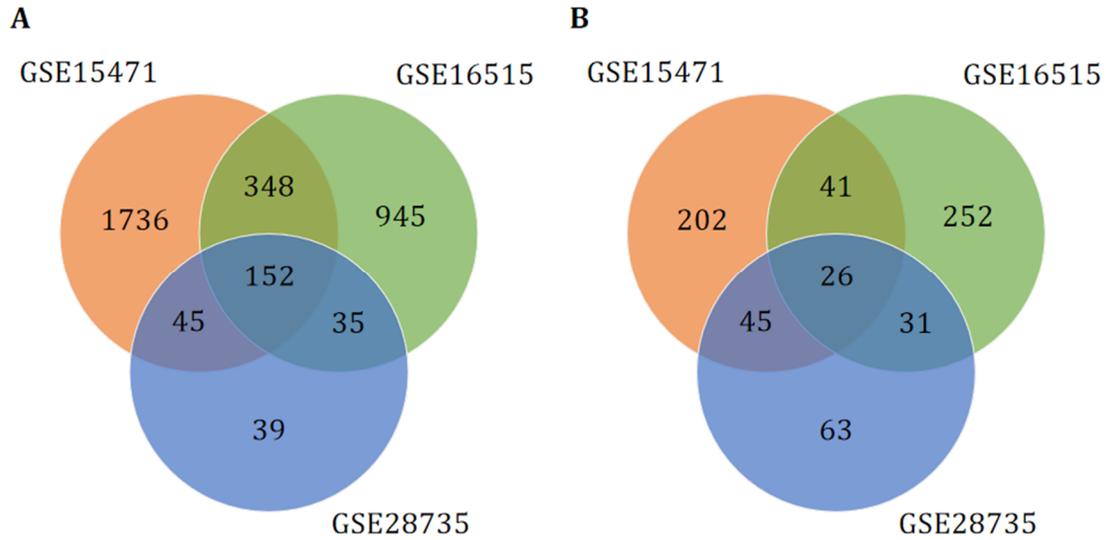

Figure 6.2: Illustration of identifying common DEGs using the Venn-Diagram

TABLE 6.5: Target genes for DEMs predicted by miRecords Tool (Numbers inside bracket of title of DEMs denotes the number of target genes predicted for that particular DEM)

| Title of DEMs | Target Genes of DEMs |
|---|---|
| Upregulated DEMs | |
| hsa-miR-135b (77) | TSEN54, GGNBP2, C6orf106, NUCKS1, TMEM168, PELI2, LZTS1, NUDT4, HOXA10, ELOVL2, ST7L, NCKIPSD, MYEF2, YBX2, RGL1, WAPAL, KIAA0831, CHSY1, BTBD10, ADO, ZRANB2, HIF1A, MTDH, PHOSPHO1, C3orf59, CHMP4B, SLC9A9, YTHDF3, ZNF385B, LONRF1, SGMS1, PHLDB2, ARHGEF7, C1orf96, KCTD12, CTTNBP2, PSIP1, JAKMIP2, TMEM97, BZW2, TAF4, STAT6, SSR2, SSR1, PPP1CC, MSX2, KPNA3, FRK, FOXO1, BACH1, ANXA7, SLC25A5, ADCYAP1, GOLGA7, PTER, NR3C2, CACNA1D, SEC62, TRPC1, ARHGAP6, TBK1, HPS5, PIM2, RALBP1, CPLX1, SMC1A, AKR1A1, MAN1A1, PRKD3, PTK2, LIMK2, ELK3, NUP153, JAK2, ENTPD4, KLF4, COL4A3 |
| hsa-miR-197 (0) | N/A |
| hsa-miR-490-3p (0) | N/A |
| hsa-miR-221 (36) | CDKN1B, ATXN1, TMCC1, NSMCE4A, GOLSYN, EIF5A2, DCUN1D1, DMRT3, C12orf30, MIER3, RBM24, NAP1L5, HECTD2, GARNL1, ASB7, INSIG1, NRK, ZEB2, RIMS3, TOX, PPP3R1, BMF, ARF4, NTF3, TCF12, EIF3J, HRB, FOS, PHF2, ARID1A, FERMT2, ZFPM2, PPARGC1A, MYLIP, VGLL4, CCDC64 |
| hsa-miR-145 (55) | MDFIC, CACHD1, EPB41L5, SEMA6A, NUFIP2, SRGAP1, RIN2, CDC37L1, KIF21A, INOC1, RAB14, YTHDF2, |
| | BACH2, ACBD3, ZBTB10, PAPD4, TMEM178, VASN, SPSB4, LENG8, SELI, SSH2, SLITRK6, ADPGK, SNX27, UXS1, TRIM2, PLCL2, NUAK1, RGS7, REV3L, PXN, LOX, DUSP6, CBFB, ACTG1, FLNB, ANGPT2, PPP3CA, SOX9, SLC1A2, ERG, AKAP12, CLINT1, PDCD4, IVNS1ABP, NEDD9, SEMA3A, CITED2, MPZL2, SCAMP3, IRS1, ABCA1, YES1, MYO5A |
| hsa-miR-10a (19) | TBX5, NARG1, HOXA3, USP46, BACH2, RAP2A, SLC38A2, ANKFY1, NCOA6, BAZ2B, ZMYND11, HAS3, GTF2H1, ELAVL2, KLF11, TFAP2C, BDNF, PAFAH1B1, SON |
| hsa-miR-484 (0) | N/A |
| hsa-miR-337-3p (0) | N/A |
| hsa-miR-342-3p (13) | PPP3R1, ZAK, BTN2A1, MRFAP1, ZIC4, RSBN1, FAM53C, PDGFRA, FUT8, UBE2D2, ID4, EDA, C1orf32 |
| hsa-miR-223 (34) | NFIA, FBXO8, STK39, SACS, RBM16, PDS5B, PHF20L1, CRIM1, FBXW7, SLC8A1, PTBP2, DERL1, C13orf18, RNF34, SLC37A3, PURB, PRDM1, MYST3, LMO2, ACVR2A, ALCAM, CBFB, F3, PCTK2, HHEX, RASA1, RPS6KB1, RHOB, EFNA1, ACSL3, PKNOX1, TSPAN7, INPP5B, MTPN |
| hsa-miR-140-3p (0) | N/A |
| hsa-miR-21 (34) | RP2, JAG1, SOX5, LEMD3, KIAA1012, BAHD1, ADNP, ASPN, CHD7, PELI1, C17orf39, MRPL9, GLCCI1, ARHGEF7, KBTBD6, PAN3, SSFA2, YAP1, TGFBI, TIMP3, PLEKHA1, MATN2, NTF3, PCSK6, PLAG1, PPP1R3A, SATB1, SKI, STAT3, NFIB, RASGRP1, SPRY1, SMAD7, PDZD2 |
| hsa-miR-331-3p (11) | B4GALT2, ZNF513, RIC8B, BAIAP2, CPSF2, CAMK2A, SNX17, PHC2, DAG1, NRP2, RIMS4 |
| hsa-miR-190b (0) | N/A |
| Downregulated DEMs | |
| hsa-miR-302c (64) | E2F7, PARP8, NEUROD6, SLC22A23, CFL2, DMTF1, TNFAIP1, ZNFX1, TSHZ3, DCUN1D1, MNT, CAMK2N1, C7orf43, ECT2, SNRK, FBXO11, DDHD1, LHX6, NAPEPLD, MTERFD2, NCOA7, CRTC2, ZNF800, C10orf104, C5orf41, YTHDF3, MIER3, PAPOLA, KBTBD8, CDCA7, SYNC1, SUV420H1, BRP44L, ZMYND11, TXNIP, NFIB, RAB11A, DUSP2, DYRK2, TFAP4, RBBP7, PPP6C, PLAG1, MEF2C, ITGB8, BCL6, SMAD2, NR4A3, RABGAP1, DERL2, LUC7L2, ZNF385A, ZDHHC17, CAMTA1, RGL1, TOX, SENP1, ZBTB11, TRPS1, ATAD2, WDR37, MKRN1, PRDM4, ESR1 |
| hsa-miR-630 (0) | N/A |

## 6.6 Prediction of microRNA Targets

The next step was to identify the target genes of the differentially expressed microRNAs (DEMs). We used an online tool, miRecords to obtain target genes of DEMs. Normally, miRecords blends the effects of more than 10 microRNA target forecast applications but no target genes were prophesied for more than 6 microRNA target forecast applications. Therefore, we accepted target genes that were prophesied by precisely 6 microRNA target forecast applications. In Table 6.5, we have displayed the target genes that were identified by miRecords.

After identifying target genes, we found out the DEGs which were also present in the target genes of selected microRNAs. And the experiment revealed that only two genes, ECT2 and NRP2, were present inside both sets: the differentially expressed genes set and the target genes set. An illustration of this phenomenon is displayed in figure 6.3 via Venn-diagram.

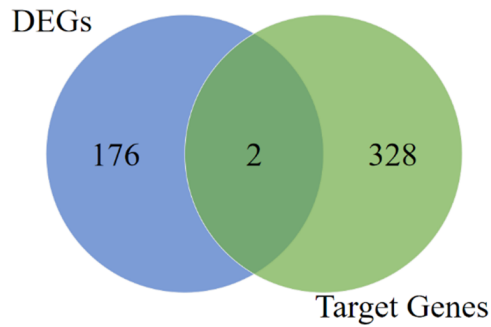

Figure 6.3: Display of the common genes among DEGs and target genes of DEMs by Venn-Diagram

## 6.7 Overall Survival Analysis

After identifying the common genes, it's time for validation of biomarkers. For this analysis, we used OncoLnc, an online tool for overall survival analysis. By creating Kaplan-Meier plots, this tool allowed us to classify victims based on the level of gene expression regarding a particular gene. Unexpected plus hazard ratios were found with 95% confidence intervals. Hence, bottom 90, 80, 60 and top 10, 20, 40 percentiles were marked as low and high groups. Overall survival analysis showed that greater mRNA expression regarding ECT2 and NRP2 was correlated with inadequate overall durability as log-rank P-value for ECT2 and NRP2 were 0.00124 and 0.0177 respectively while lower and upper percentiles were set to 10% and 90% respectively. Figure 6.4 shows the Kaplan-Meier plots for 10-90 lower-upper percentile for ECT2 and NRP2 respectively.

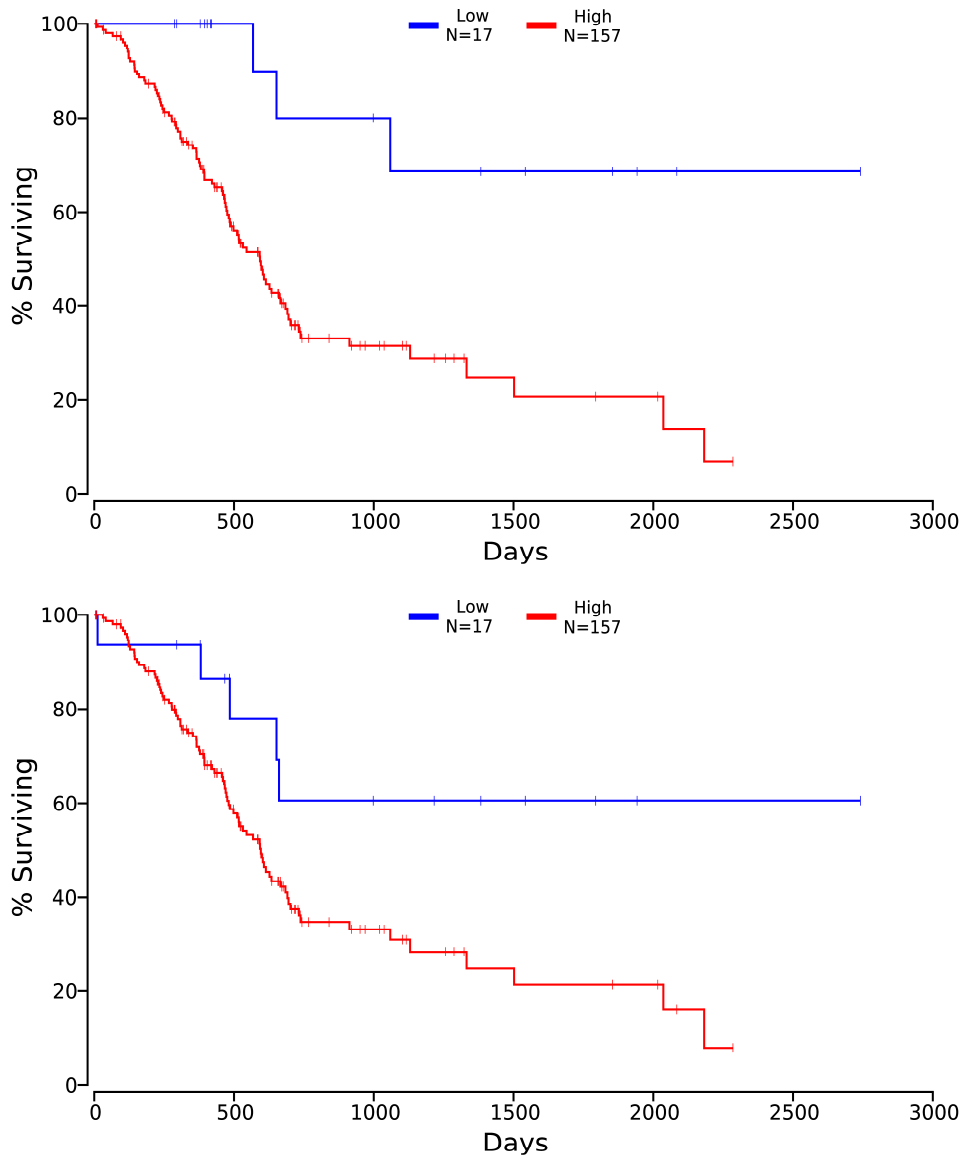

Figure 6.4: Presentation of ECT2 (top) & NRP2 (bottom) for 10-90 percentile with log-rank p-value 0.00124 and 0.0177 respectively

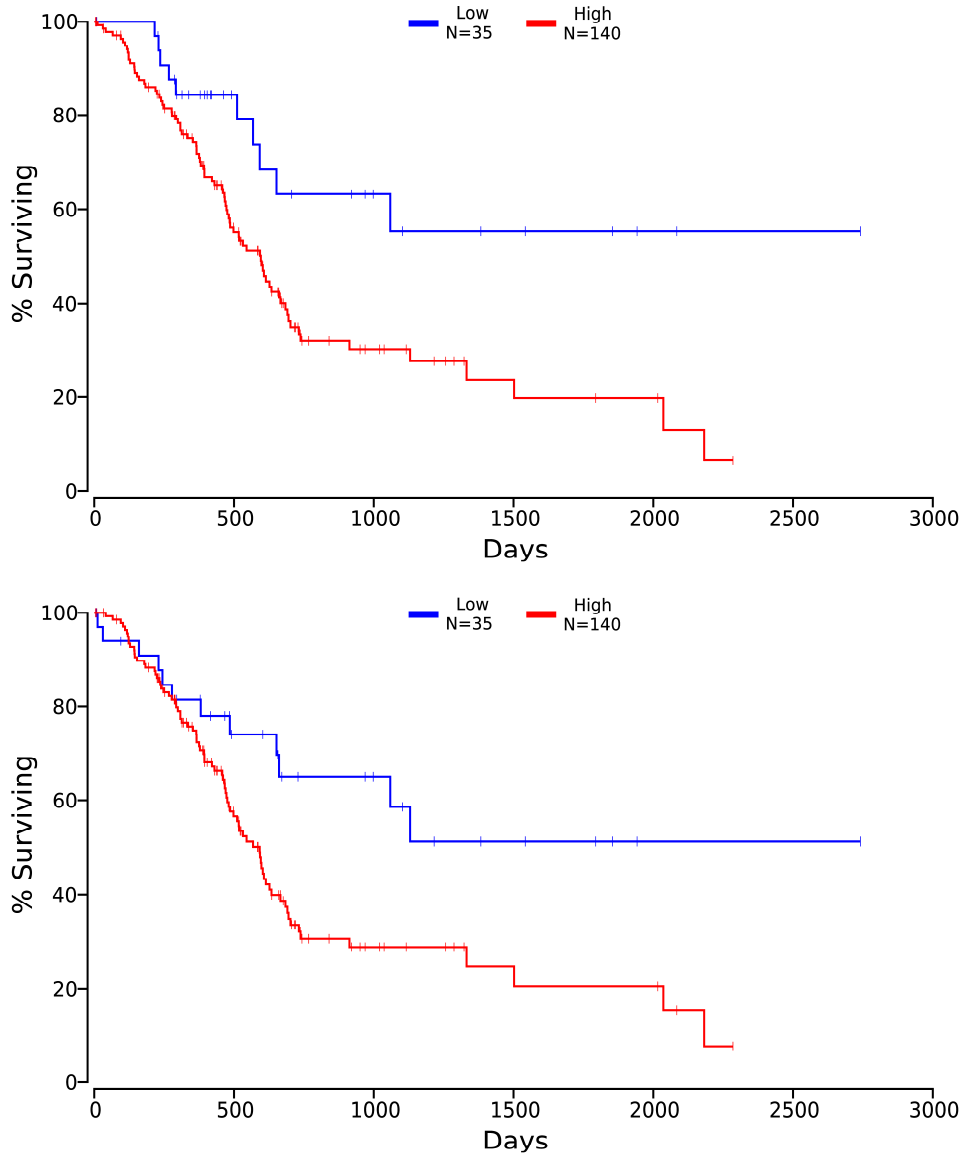

Figure 6.5: Presentation of ECT2 (top) & NRP2 (bottom) for 20-80 percentile with log-rank p-value 0.00484 and 0.00801 respectively

After that, the percentage was changed from 10% vs 90% to 20% vs 80% and 40% vs 60%. For the percentage of 20% vs 80%, obtained log-rank P values were 0.00484 and 0.00801 and for the percentage of 40% vs 60%, obtained log-rank P values were 0.000851 and 0.059 for ECT2 and NRP2 respectively. Figure 6.5 and 6.6 illustrates these phenomena.

The percentile table used for the overall survival analysis is available publicly in OncoLnc site: www.oncolnc.org. The data has four main columns: patient no, days alive, alive or dead and expression of ECT2 or NRP2. For both tables, the data is displayed in a sorted arrangement from lower to higher with respect to expression rate of ECT2 or NRP2 to define low and high group.

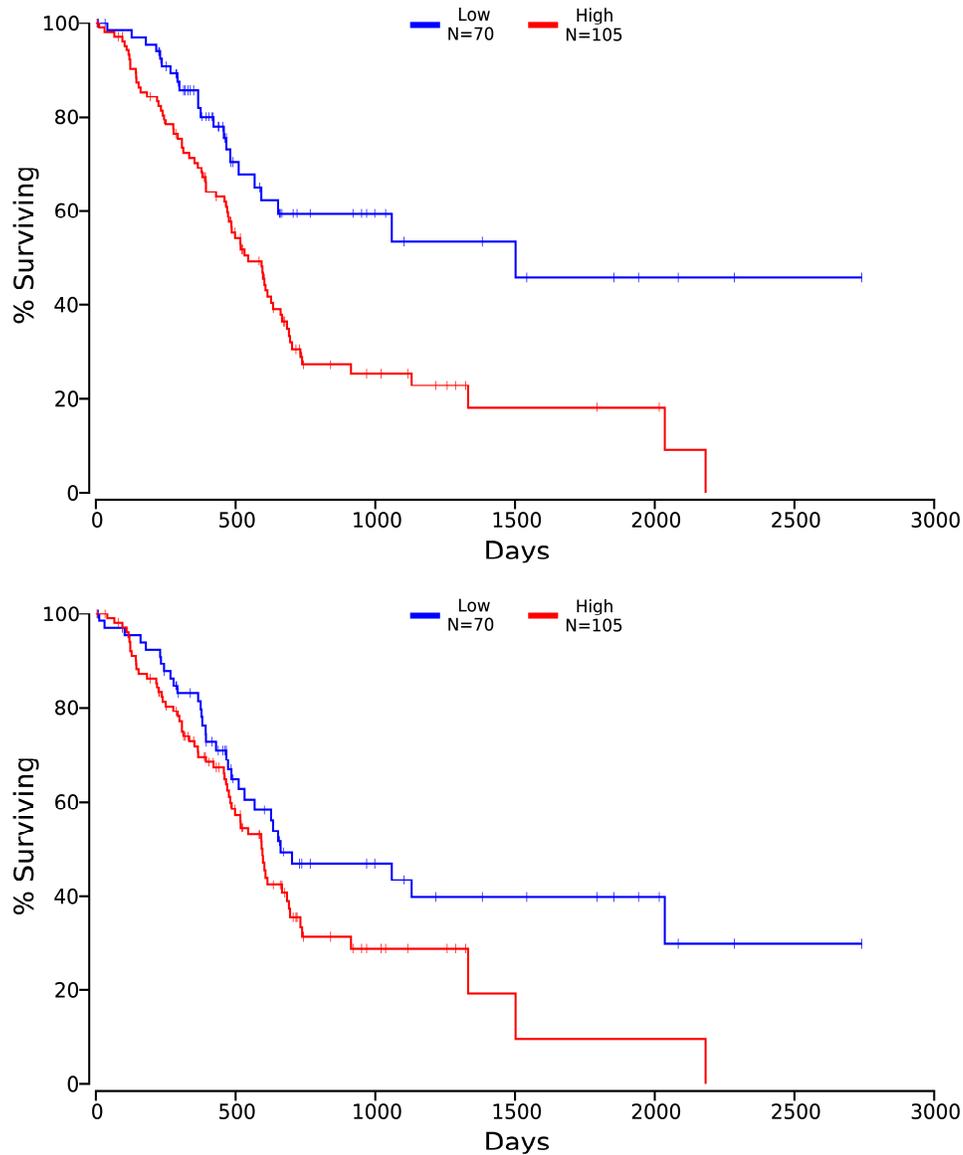

Figure 6.6: Presentation of ECT2 (top) & NRP2 (bottom) for 40-60 percentile with log-rank p-value 0.000851 and 0.059 respectively

Moreover, associations between other popular cancers for the selected DEGs were obtained and investigated (Table 6.6 and Table 6.7). The investigation showed that ECT2 and NRP2 were actually connected to worse overall durability particularly for particularly pancreatic cancer based on the log-rank P-values which allowed us to come to a conclusion that ECT2 and NRP2 may act as therapy target or potential prognostic biomarkers for pancreatic cancer.

TABLE 6.6: Comparison of log-rank P-values for different cancers for different percentiles for ECT2

| Cancer Title | 10-90 Percentile | 20-80 Percentile | 40-60 Percentile |
|---|---|---|---|
| BLCA | 0.487 | 0.455 | 0.55 |
| BRCA | 0.956 | 0.188 | 0.405 |
| CESC | 0.192 | 0.838 | 0.326 |
| COAD | 0.963 | 0.925 | 0.0683 |
| ESCA | 0.736 | 0.751 | 0.905 |
| GBB | 0.248 | 0.879 | 0.0405 |
| HNSC | 0.0141 | 0.0154 | 0.0135 |
| KIRC | 0.365 | 0.486 | 0.543 |
| KIRP | 0.355 | 0.38 | 0.206 |
| LAML | 0.101 | 0.0135 | 0.125 |
| LGG | 0.0227 | 0.000548 | 0.00000225 |
| LIHC | 0.00295 | 0.00362 | 0.00157 |
| LUAD | 0.0433 | 0.0308 | 0.00000784 |
| LUSC | 0.00772 | 0.127 | 0.163 |
| OV | 0.193 | 0.997 | 0.392 |
| PAAD | 0.00124 | 0.00484 | 0.000851 |
| READ | 0.378 | 0.0501 | 0.0648 |
| SARC | 0.0921 | 0.0624 | 0.137 |
| SKCM | 0.00126 | 0.0114 | 0.163 |
| STAD | 0.983 | 0.295 | 0.0952 |
| UCEC | 0.019 | 0.0427 | 0.0169 |

In Table 6.6 and 6.7, the short forms have been used to denote each cancer type. The full form of these cancers are - BLCA: Bladder Urothelial Carcinoma, BRCA: Breast invasive carcinoma, CESC: Cervical squamous cell carcinoma and endocervical adenocarcinoma, COAD: Colon adenocarcinoma, ESCA: Esophageal carcinoma, GBM: Glioblastoma multiforme, HNSC: Head and Neck squamous cell carcinoma, KIRC: Kidney renal clear cell carcinoma, KIRP: Kidney renal papillary cell carcinoma, LAML:

Acute Myeloid Leukemia, LGG: Brain Lower Grade Glioma, LIHC: Liver hepatocellular carcinoma, LUAD: Lung adenocarcinoma, LUSC: Lung squamous cell carcinoma, OV: Ovarian serous cystadenocarcinoma, PAAD: Pancreatic adenocarcinoma, READ: Rectum adenocarcinoma, SARC: Sarcoma, SKCM: Skin Cutaneous Melanoma, STAD: Stomach adenocarcinoma, UCEC: Uterine Corpus Endometrial Carcinoma.

TABLE 6.7: Comparison of log-rank P-values for different cancers for different percentiles for NRP2

| Cancer Title | 10-90 Percentile | 20-80 Percentile | 40-60 Percentile |
|---|---|---|---|
| BLCA | 0.0777 | 0.142 | 0.0127 |
| BRCA | 0.57 | 0.947 | 0.951 |
| CESC | 0.613 | 0.953 | 0.72 |
| COAD | 0.527 | 0.315 | 0.778 |
| ESCA | 0.939 | 0.93 | 0.618 |
| GBB | 0.989 | 0.232 | 0.978 |
| HNSC | 0.751 | 0.218 | 0.0931 |
| KIRC | 0.374 | 0.303 | 0.526 |
| KIRP | 0.529 | 0.226 | 0.0305 |
| LAML | 0.279 | 0.187 | 0.0826 |
| LGG | 0.00879 | 0.142 | 0.828 |
| LIHC | 0.00763 | 0.163 | 0.0613 |
| LUAD | 0.975 | 0.489 | 0.619 |
| LUSC | 0.252 | 0.664 | 0.63 |
| OV | 0.992 | 0.892 | 0.585 |
| PAAD | 0.0177 | 0.00801 | 0.059 |
| READ | 0.828 | 0.55 | 0.928 |
| SARC | 0.599 | 0.958 | 0.416 |
| SKCM | 0.766 | 0.3 | 0.773 |
| STAD | 0.091 | 0.0134 | 0.137 |
| UCEC | 0.052 | 0.529 | 0.873 |

## 6.8 Comparative Study

Table 6.8 presents a comparative study with the previous research [39]. Here, we have presented the comparison on the basis of methods used for selecting differentially expressed genes and microRNAs, number of DEGs and DEMs selected, number of target genes selected and number of probabilistic prognostic biomarkers identified for each research.

TABLE 6.8: Comparison between our research with previous study

| Criteria | Previous Research | Our Research |
|---|---|---|
| Number of Common DEGs | 236 | 178 |
| Number of DEMs | 21 | 16 |
| Number of Target Genes | 512 | 330 |
| Identified Biomarkers | ECT2, NR5A2, NRP2, TGFBI | ECT2, NRP2 |

## 6.9 Conclusion

In this chapter, we have discussed and presented our overall experimental analysis with all required tables and figures. We basically divided our experiment into three broad sections: identification of differentially expressed genes (DEGs) and microRNAs (DEMs), identification of target genes by miRecords tool and overall survival analysis by OncoLnc tool. The outcome was two probabilistic biomarkers, ECT2 and NRP2, for the prognosis of pancreatic cancer.

# CHAPTER 7
# Conclusion and Future Scopes

*Introduction*

*Conclusion*

*Future Scopes*

## 7.1 Introduction

This chapter summarizes the whole research in several words describing the problem domain, previous researches, our contribution, experimental analysis and the outcomes of this research. Moreover, we have provided a brief study about the future scopes of this research.

## 7.2 Conclusion

Cancer-related death has been a research-interest for decades now as cancer is one the biggest cause of death among humans. Among all the cancer types, pancreatic cancer is the most lethal one having the worst 5-year survival rate of less than 5% only. Statistics have revealed that most of the pancreatic cancer victims are diagnosed in the advanced stage and because of the lack of diagnostic treatments the survival rate of pancreatic cancer is extremely low [6-7]. Early diagnosis is the only way to improve the survival rate for pancreas patients [6-7]. Recent studies suggested four prognostic biomarkers for pancreas using online tools [39]. However, in this research, we revealed that statistical methods can provide better and fast results. We downloaded three mRNA microarray and one microRNA expression dataset for GEO. The datasets were then fed to some well-established feature selection methods (Student's T-Test, Kruskal-Wallis Test and Wilcoxon-Mann-Whitney Test) to find out the differentially expressed genes (DEGs) and differentially expressed microRNAs (DEMs). We used an online tool, miRecords, to identify the target genes of the DEMs. These target genes were then intersected with the common DEGs of three mRNA datasets. Two genes that were present in target genes set and DEGs set, ECT2 and NRP2, were identified as the prognostic biomarkers of pancreatic cancer. The overall survival analysis reported that the survival rate is low for high expression of ECT2 and NRP2 and high for low expression of the genes. Furthermore, we compared the log-rank p-value among different kinds of cancers to present this fact. Finally, we compared the outcome of our research with previous studies and concluded that ECT2 and NRP2 may play a significant role as prognostic biomarkers in pancreatic cancer.

## 7.2.1 Epithelial Cell Transforming 2 (ECT2)

The protein encoded by this gene is a guanine nucleotide exchange factor and reconstructing protein that is associated with Rho-specific exchange factors and yeast cell cycle regulators. The expression of this gene is raised with the onset of DNA synthesis and remains elevated during G2 and M phases. In situ hybridization analysis revealed that expression is at a high level in cells undergoing mitosis in regenerating liver. Thus, this protein is manifested in a cell cycle-dependent manner during liver regeneration and is believed to possess an essential part in the regulation of cytokinesis. Several transcript variants encoding various isoforms have been observed for this gene (provided by RefSeq, Mar 2017) [179].

## 7.2.2 Neuropilin 2 (NRP2)

This gene encodes a member of the neuropilin family of receptor proteins. The encoded transmembrane protein ties to SEMA3C protein {sema domain, immunoglobulin domain (Ig), short basic domain, secreted, (semaphorin) 3C} and SEMA3F protein {sema domain, immunoglobulin domain (Ig), short basic domain, secreted, (semaphorin) 3F}, and communicates with vascular endothelial growth factor (VEGF). This protein may perform a function in cardiovascular improvement, axon guidance, and tumorigenesis. Multiple transcript variants encoding distinct isoforms have been recognized for this gene (provided by RefSeq, Jul 2008]) [179].

## 7.3 Future Scopes

We believe that our research will have a significant impact on the development of therapy target for pancreatic cancer. However, we acknowledge that there is much room for future works in this domain. The research community is being involved in microarray data analysis more and more as time flies. In our research, we have studies four well established and popular datasets. Nevertheless, more datasets are available regarding microarray analysis of pancreatic cancer tissue. More datasets will be introduced in the near future as well. These datasets may contain much more information (genes) than the available datasets. The same scenario is applied for microRNA expression dataset as well. Analyzing more sophisticated datasets may reveal more prognostic biomarkers for pancreatic cancer as well. Moreover, the scientific community is moving towards next-generation sequencing for gene expression analysis as a next-generation sequence provides more accurate results due to the small loss in information compared to microarray technology. This can also be a significant field of interest. The main scope of future work for prognostic biomarker identification is, however, validating the pathway via real-time lab-experiments. Casper Cas9 gene-editing system can be used on real-time pancreas tissue to validate if the biomarkers can be used as prime therapy target that will eventually improve the survival rate. We believe more researchers will be attracted to this domain and more advancements will be introduced in the near future. The following sections will discuss the two prognostic biomarkers.

# LIST OF PUBLICATIONS

**Publication No. 01**

**Publication Title:** Prognostic Biomarker Identification for Pancreatic Cancer by Analyzing Multiple mRNA and microRNA Datasets

**Authors:** Azmain Yakin Srizon, Md. Al Mehedi Hasan

**Conference:** 5th International Conference on Computer, Communication, Chemical, Materials & Electronic Engineering (IC4ME2 -2019)

**Organized By:** Faculty of Engineering, University of Rajshahi

**Venue:** University of Rajshahi, **Date:** 11-12 July 2019

**Abstract:** Having the five-year survival rate of approximately 5%, currently, the fourth leading reason for cancer-related deaths is pancreatic cancer. Previously, various works have concluded that early diagnosis plays a significant role in improving the survival rate and different online tools have been used to identify prognostic biomarker which is a long process. We think that the statistical feature selection method can provide a better and faster result here. To establish our statement, we selected three different mRNA microarray (GSE15471, GSE28735 and GSE16515) and a microRNA (GSE41372) dataset for identification of differentially expressed genes (DEGs) and differentially expressed microRNAs (DEMs). By using a parametric test (Student's t-test), 178 DEGs and 16 DEMs were selected. After identifying target genes of DEMs, we selected two DEGs (ECT2 and NRP2) which were also identified among DEMs target genes. Furthermore, overall survival analysis confirmed that ECT2 and NRP2 were correlated with inadequate overall survival. Hence, we concluded that for pancreatic cancer, a parametric test like Student's t-test can perform better for biomarker identification, and here, ECT2 and NRP2 can act as possible biomarkers. All the resources, programs and snippets of our literature can be discovered at https://github.com/Srizon143005/PancreaticCancerBiomarkers.